\documentclass[%
 reprint,
showpacs,
 amsmath,amssymb,
 aps,
]{revtex4-1}

\pdfoutput=1
\usepackage[pdftex]{graphicx}
\usepackage{dcolumn}
\usepackage{bm}

\newcommand{\pa}[2]{\frac{\partial{#1}}{\partial{#2}} } 
\newcommand{\pazwei}[2]{\frac{\partial^2{#1}}{\partial{#2^2}} } 
  
\newcommand{\D}{\mathrm{d}}
\newcommand{\mw}[1]{\langle #1 \rangle}

\newcommand{\vq}{\vec{q}}
\newcommand{\mph}{\bar{\phi}}

\renewcommand{\vec}[1]{\boldsymbol{#1}}

\begin{document}


\title{Protocol-independent granular temperature supported by numerical
simulations}

\author{Volker Becker}
  \email{Volker.Becker@ovgu.de}
\author{Klaus Kassner}%
 \email{Klaus.Kassner@ovgu.de}
\affiliation{%
  Institute for Theoretical Physics, Otto von Guericke University Magdeburg\\
  Postfach 4120, 39106 Magdeburg, Germany
}%




\date{August 27, 2015}

\begin{abstract}
  A possible approach to the statistical description of granular
  assemblies starts from Edwards' assumption that all blocked states
  occupying the same volume are equally probable (S.F. Edwards, R.
  Oakeshott, Physica A \textbf{157}, 1080 (1989)). We performed
  computer simulations using two-dimensional polygonal particles
  excited periodically according to two different protocols:
  excitation by pulses of ``negative gravity'' and excitation by
  ``rotating gravity''. The first protocol exhibits a non-monotonous
  dependency of the mean volume fraction on the pulse strength. The
  overlapping histogram method is used in order to test whether or not
  the volume distribution is described by a Boltzmann-like
  distribution, and to calculate the inverse compactivity as well as
  the logarithm of the partition sum. We find that the mean volume is
  a unique function of the measured granular temperature,
  independently of the protocol and of the branch in $\phi(g)$, and
  that all determined  quantities are in agreement with Edwards'
  theory.
\end{abstract}

\maketitle


\section{Introduction}

Granular materials are typically composed of thousands to millions of
individual particles (or more). Think for example of the cereals for
breakfast or of the sand on the shore. These large numbers suggest
that statistical methods may be applicable and constitute a powerful
tool in developing a better theoretical understanding of these kinds
of materials.  However, contrary to the situation in gases or fluids,
where the permitted phase space is explored continuously due to
chaotic molecular motion, thermal fluctuations are negligible for
granular materials.  Furthermore, the particle dynamics are
dissipative.  Therefore, it is not possible to carry standard
statistical mechanics over to granular assemblies. In particular, the
classical Boltzmann distribution, where the probability of a state is
inversely proportional to the exponential of its energy (measured in
units of $k_B T$), will not apply to these systems.

Edwards and Oakeshott \cite{Edwards1989,Edwards1989a} proposed that
concepts from classical statistical mechanics are applicable, if one
assumes that the volume of a static, stable granulate plays the same
role in granular statistics as the energy of a microstate in classical
statistics. This means that, analogous to the classical microcanonical
ensemble, where all states with the same energy are equally probable,
all mechanically stable configurations of the granular assembly that
occupy the same volume occur with the same probability.  The entropy
of this granular microcanonical ensemble is proportional to the
logarithm of the number of blocked states with a certain volume.  By
analogy with classical statistics, one can define a temperature-like
variable, called compactivity, as the inverse of the derivative of the
Edwards entropy with respect to the volume.  Later, it turned out that
for a full description of a granular system, beyond the volume
ensemble also an ensemble for the different stress states must be
introduced \cite{Bi2015,Henkes2007,Blumenfeld2006} and that the volume
and the stress ensembles are probably interdependent
\cite{Blumenfeld2012}. However, it seems that expectation values of
quantities that depend on the geometrical state only and not on the
stress state are well described by the volume ensemble
\cite{Blumenfeld2014}, possibly because the force-moment tensor can be
treated as approximately constant in the systems considered here.
Therefore we focus on the volume ensemble in most of the present
paper.

Some doubts about the temperature-like interpretation of the
compactivity have been raised on the basis of equilibration
experiments performed by Puckett and Daniels \cite{Puckett2013}.  They
found that in a two-component system, the compactivity did not
equilibrate, whereas the angoricity did. There are different ways to
explain this observation, some of which do not necessitate to give up
the interpretation of the compactivity as a temperature-like variable
\footnote{If we assume exchange of the volume associated with a
  particle to be restricted between the two particle types, an
  incomplete equilibrium with different compactivities might arise. A
  thermodynamic analogue would be the fountain effect, where two
  receptacles with superfluid helium are connected by a
  supercapillary, allowing only particle exchange of the superfluid
  component, but no energy exchange. Due to the restricted energy
  exchange, equilibrium configurations with different temperatures
  (and pressures) in the two containers exist. Alternatively, it is
  possible that this was a protocol related effect or even a finite
  size effect, because the subsystem consisted only of 100
  particles.}.

A key assumption of standard statistical mechanics is the equivalence of
time and ensemble averages, but mechanically stable granular
configurations are static states without any intrinsic time evolution.
On the other hand, by applying the same external excitation to
the granular material again and again (i.e.~tapping \cite{Nowak1998}
or shearing \cite{Tsai2003}), this external excitation may take over
the role of thermal agitation and the concept of a time average becomes
meaningful for granular statistics as well. Some tests of the ergodicity
of granular systems are available in the literature. With systems of frictional
discs excited with flow pulses, equivalence between time and ensemble
averages was found \cite{Ciamarra2006}. In the case of a vertical tapping
protocol, dependency on the protocol parameters was noted. Especially
for small tapping amplitudes, nonergodicity was observed in 
numerical simulations \cite{Paillusson2012}. This may be related to
the occurrence of irreversible branches in tapped granular systems
\cite{Nowak1998,Pugnaloni2008}. 

Several methods were proposed to determine the compactivity from
experimental or simulation data
\cite{Nowak1998,Schroeter2005,Zhao2012,Zhao2014} and applied to
different kinds of granulates. In this work, we employ two-dimensional
discrete-element (DEM) simulations using polygonal particles to apply
two different excitation protocols to otherwise identical granulates.
This allows us to determine whether the calculated compactivity is
independent of the specific excitation protocol (which would support
the Edwards theory) or not (which would oppose it). Recently, some
authors have addressed the issue whether or not it is necessary to
introduce protocol specific extensions of Edwards-like approaches
\cite{Paillusson2015,Asenjo2014}.  We briefly discuss a possible
approach in section \ref{sec:protocoldependence}.  At least for the
two protocols considered in this paper, we find protocol independence.
Furthermore, we test whether ideal-gas-like prediction analogies of
the Edwards theory are consistent with the simulation data, and find
good qualitative agreement, which even becomes quantitative if certain
parameters are chosen by fitting.
 
The paper is organized as follows. In section \ref{sec:theo}, we give
an overview of some aspects of Edwards' theory, relevant to this work.
In section \ref{sec:method}, we introduce the simulation technique.
Section \ref{sec:protocol} gives details of the applied excitation
protocols and in section \ref{sec:anwendung}, the simulation results
will be presented, evaluated and discussed. In Sec.~\ref{sec:limits},
some limitations of the volume ensemble are illustrated.  Finally,
section \ref{sec:conclusion} gives a summary of our findings.


\section{Theoretical framework}\label{sec:theo}
\subsection{The microcanonical and the canonical volume ensemble}

The main assumption of the Edwards theory is the
following \cite{Edwards1989,Edwards1989a}: If a granular ensemble is
generated by a {\em reproducible} preparation protocol, all resulting
mechanically stable configurations of the granular system which
occupy the same volume will occur with the same probability on
repetition of the protocol.
This means that in Edwards' granular statistics the volume plays the
same role as the energy in ordinary statistics.  Consequently, the
entropy $S$ of the microcanonical ensemble having a certain volume $V$
is given as the logarithm of the number $\Omega$ of stable states (in
the permitted phase space) occupying this volume.

Let a microstate of the granular ensemble, comprised of $N$ particles,
be described by a set of variables $\vec{q}$. The entropy of this
state is given by (see e.~g.
\cite{Edwards1989,Edwards1989a,Blumenfeld2014}):
\begin{align} 
      S(V,N)      &= \ln \Omega(V,N)\>,  \label{eq:entropiedefinition}  \\
      \Omega(V,N) &= \int_{\{\vec{q}\}} \D \vec{q} ~ \delta(V-W(\vec{q}))\>.
\end{align}
The integral runs over all stable configurations $\{\vec{q}\}$ and the
function $W(\vec{q})$ gives the volume of the state $\vec{q}$. Note
that in this context, $W(\vq)$ is the granular analogue of the
Hamiltonian and $V$ is the analogue of the internal energy.

By analogy with standard statistical mechanics, an intensive,
temperature-like variable $\chi$ can be defined, usually called
compactivity:
\begin{equation}
   \chi = \beta^{-1} = \pa{V}{S}\>.  \label{eq:compdefinition}
\end{equation}
In this paper, we will, for simplicity, mostly use the
``thermodynamic beta'', defined as the inverse compactivity
$\beta=1/\chi$, instead of the compactivity itself.

Note that in general, a constant $\lambda$ analogous to
the Boltzmann constant may be introduced in \eqref{eq:entropiedefinition} and 
\eqref{eq:compdefinition}. Here, we use $\lambda=1$ which means that we measure
compactivity in units of volume.  

As in ordinary statistics/thermodynamics (see e.g. \cite{Landau1987})
one may switch from the microcanonical ensemble to the canonical
ensemble by a Legendre transformation. In the corresponding canonical
ensemble, the probability of a microstate $\vq$ occupying the
volume $W(\vq)$ is given by a Boltzmann-like distribution:
\begin{equation}
   P(\vq)=\frac{1}{Z} e^{- \beta W(\vq)} \label{eq:boltzmanlike}
\end{equation}
with the canonical partition function \cite{Lechenault2006,Blumenfeld2012,Zhao2014}
\begin{equation}
  Z= \int_{\{q\}} \D \vq ~ e^{- \beta W(\vq)}\>. \label{eq:kanonischeZustandssumme}
\end{equation}
Once again, the integral runs over all possible mechanically stable
states. Note that this is equivalent to a notation often used in
the literature, where the integral runs over all states and contains an
additional factor $\theta(\vq)$, which takes the value zero for
forbidden states and one for allowed ones, thus selecting the
permitted states.

Note that a tapping protocol does not necessarily lead to canonical
sampling of the system. A trivial example would be tapping with so
small an amplitude that the system is trapped in the initial blocked
state. A non-trivial example is the irreversible branch for vertical
tapping observed by Nowak \emph{et.~al.}~\cite{Nowak1998}. Also, the
comparison between the analytically solvable Bowles-Ashwin model
system \cite{Bowles2011} and simulations where such a system is
vertically tapped \cite{Irastorza12013} exhibits deviations from the
canonical ensemble prediction. Note that the Bowles-Ashwin model
assumes a highly confined geometry with much lower complexity than
realistic granular systems. Vertical tapping applied to
this special system with strong confinement may be unsuited for phase
space exploration according to a flat probability measure. Or else the
flat Edwards measure does not hold in general.  Even if that were the
case, a meaningful definition of compactivity might still be possible
as will be described now.

\subsection{Generalised (protocol dependent) ensembles}
\label{sec:protocoldependence}
In case it turned out that Edwards' assumption of a flat
probability measure is not satisfied for the microcanical ensemble, it
is possible to modify the ensemble with (in general, state and
protocol dependent) weighting factors $w(\vq)$
\cite{Asenjo2014,Bi2015,Paillusson2015}.  The microcanical state
density would be modified as follows
\begin{equation} 
      \Omega_{G}(V,N) = \int_{\{\vec{q}\}} \D \vec{q} ~ w(\vq)   \delta(V-W(\vec{q}))\>. \label{eq:generalisedmicrocanic}
\end{equation}
Providing we have 
\begin{equation} 
\Omega_{G}(V_1+V_2,N) = \Omega_{G}(V_1,N_1)\,\Omega_{G}(V_2,N_2)\>,
\end{equation}
which is a much weaker assumption than a flat probability measure
\cite{Bi2015}, it is still possible to define a compactivity.  The
canonical formulation for the volume ensemble then reads
\begin{align}
   P(\vq) &= w(\vq) \frac{1}{Z} e^{- \beta W(\vq)} \>,
\label{eq:generalisedBoltzman}\\
   Z_{G} &= \int_{\{q\}}  \D \vq ~ w(\vq) e^{- \beta W(\vq)}\>.   \label{eq:generalisedParti}
\end{align}
A similar approach is feasible for the force-moment and the combined
ensembles.  Generalised canonical ensembles of this kind are 
used in statistical genetics \cite{Nourmohammad2013,Barton2009}.

The qualitative impact of these modifications on the resulting
thermodynamics is small. Especially equations
\eqref{eq:volflukbetarelation} and \eqref{eq:overlappinghistofit}
remain unaffected, as long as the weighting factors are the same for
all data samples. If only one protocol is used, it is impossible to
detect protocol dependencies in the weighting factors. But a
comparison of different protocols, applied to the same system, may
help to decide whether these factors, if their introduction should
turn out necessary, are protocol independent or not.

\subsection{Mean volume and volume fluctuations}

The calculation of the mean volume and its fluctuations is
straightforward.  The first derivative of the logarithm of
\eqref{eq:kanonischeZustandssumme} (or \eqref{eq:generalisedParti})
with respect to $\beta$ essentially is the mean volume
\begin{equation}
  \mw{V}=-\pa{}{\beta} \ln Z = \frac{1}{Z}  \int_{\{q\}} \D \vq ~ W(\vq)
  e^{- \beta W(\vq)}\>, \label{eq:meanVolume}
\end{equation}
whereas its second derivative is the variance of the volume
distribution, i.e., a measure for the strength of fluctuations.
\begin{align}
  \sigma_V^2 &= \mw{V^2} -\mw{V}^2=  \pazwei{}{\beta} \ln Z\>,
 \label{eq:volumefluctuations1} \\
  \sigma_V^2 &= -\pa{}{\beta} \mw{V}\>. \label{eq:sigmaVvonv}
\end{align}
Instead of the volume itself, we use the the volume fraction $\phi$,
defined as 
the sum of the grain volumes $V_g$ divided by the volume $W$ occupied
by the granulate:
\begin{equation}
    \phi(\vq)=\frac{V_g}{W(\vq)}\>.
\end{equation}
The volume $W(\vq)$ of a certain state can be written as a sum of the
mean volume $V=\mw{W}$ and the deviation from the mean: $W=V +
\Delta V$. Under the assumption that the volume fluctuations are small
compared to the mean volume, we can write:
\begin{equation}
  \phi=\frac{V_g}{V + \Delta V} \simeq
  \frac{V_g}{V}-\frac{V_g}{V^2} \Delta V \>. \label{eq:phisim}
\end{equation}
The mean volume fraction is therefore:
\begin{equation}
 \mph= \mw \phi = \frac{V_g}{V}\>, \label{eq:phimw}
\end{equation}
and from eqs.~(\ref{eq:phisim},\ref{eq:phimw}), together with
$\mw{\Delta V^2}=\sigma_V^2$, we obtain
\begin{equation}
 \sigma^2_\phi = \mw{(\phi-\mph)^2}=\frac{\mw{\phi}^4}{V_g^2} \sigma_V^2\>. 
\end{equation}
Substituting $\sigma_V^2$ from eq.~\eqref{eq:sigmaVvonv} and
reexpressing the differential of $\mw{V}$ according to
$d\mw{V}=-V_g/\mw{V}^2 d \mph$, we find a relation between the mean
volume fraction and its fluctuations:
\begin{equation}
 \sigma_\phi^2 =  \frac{\mph^2}{V_g} \pa{\bar{\phi}}{\beta}\>. 
\label{eq:volflukbetarelation}
\end{equation}
Measuring the volume fraction fluctuations as a function of the mean
volume fraction and integrating equation
\eqref{eq:volflukbetarelation} or \eqref{eq:sigmaVvonv}, respectively,
is a way to calculate the compactivity up to an unknown constant. It is used
frequently, e.g., in \cite{Nowak1998,Schroeter2005,Zhao2014}. Note
that determining the granular temperature via
eq.~\eqref{eq:sigmaVvonv} or eq.~\eqref{eq:volflukbetarelation} is
just a rule of calculation, provided by Edwards' theory but no proof
of the theory.  On the other hand, if after determining $\beta$ in
some different way the relationship \eqref{eq:volflukbetarelation}
linking it with the volume fraction were not satisfied, a
contradiction to Edwards' theory would have been demonstrated.


\subsection{Overlapping histogram method}

Another way to determine the inverse compactivity is the overlapping
histogram method proposed in 2003 by Dean and Lef\`evre
\cite{Dean2003}. This method may also be used as a test whether or not
a distribution of blocked states is Boltzmann-like distributed. In the
original paper, the method was applied to the energy of the
Sherrington-Kirkpatrick model for spin glasses, driven by a
tapping-like mechanism, which has some similarities to granular
dynamics. The method was then frequently used to determine the
granular temperature
\cite{Zhao2014,McNamara2009,Ciamarra2012,Puckett2013}.  Under the
assumption that Edwards' theory holds, the probability to measure a
certain volume V in a granulate with an inverse compactivity $\beta_0$
and a fixed number of grains $N$ is
\begin{align}
  P(V,\beta_0,N)&= \int_{\{\vq\}} \D \vq ~ \delta(V-W(\vq))
  \frac{1}{Z(\beta_0,N)} e^{- \beta_0 W(\vq)}
  \nonumber  \\
  &= \frac{\mathfrak{D}(V,N)}{Z(\beta_0,N)} e^{-\beta_0 V}\>,
  \label{eq:probfromvol}
\end{align}
where $\mathfrak{D}(V,N)$ is the number of blocked states with volume
$V$, i.e $\mathfrak{D}(V,N)=\Omega(V,N)$. This equation holds even for
the generalised non-flat probability measure approach, when eq.
\eqref{eq:generalisedmicrocanic} is used for $\Omega(V,N)$.

We define the quantity $Q$ as the logarithm of the ratio of
probability densities for the volume $V$ to arise, on the one hand at an
inverse compactivity $\beta_1$ and on the other hand, in a reference
system held at a different inverse compactivity $\beta_0$:
\begin{align}
  Q &:= \ln \frac{P(V,\beta_1,N)}{P(V,\beta_0,N)}\>,
  \label{eq:Qdefinition}
  \\
  Q &= \underbrace{(\beta_0-\beta_1)}_{:=A_{10}} V + \underbrace{\ln
    \frac{Z(\beta_0,N)}{Z(\beta_1,N)}}_{:=B_{10}}
  \label{eq:overlappinghistofit}
\end{align}
Therefore, if we measure the probability density $P(V,\beta,N)$ for
different compactivities and calculate $Q(V)$, the resulting function
must be linear if Edwards' theory holds.  Evaluating the slope
$A_{10}$ allows us to determine the inverse compactivity up to an
additive constant. Furthermore, the intercept $B_{10}$ is nothing else
than the logarithm of the partition function up to an additive
constant at the corresponding inverse compactivity.  This permits
testing the validity of Edwards' assumption in the sense, that if
$Q(V)$ were not a straight line, neither Eq.~\eqref{eq:boltzmanlike}
nor Eq.~\eqref{eq:generalisedBoltzman} would describe the probability
density of the system correctly.

However, one has to  be somewhat careful with the interpretation of the results,
as has been pointed out in \cite{McNamara2009}. Under certain
circumstances, very similar results as the ones expected from Edwards'
theory can occur, if the distributions of the samples are just
Gaussian. In appendix \ref{app:gauss}, we discuss this situation.



\subsection{Ideal quadron solution}

There are very few real ab initio predictions from Edwards' theory in
the literature
\cite{Blumenfeld2003,Blumenfeld2012,Lechenault2006,Aste2008}, due to
some general difficulties. In order to calculate analytical
expressions for the partition function, knowledge of an explicit
expression for the granular Hamiltonian $W(\vec{q})$ is necessary. Of
course, if all positions, orientations and shapes of the grains are
known, the occupied volume is a function of these quantities, but in
practice it is not easy to write down an explicit equation and even if
this can be done, the integration over the permitted blocked states is
very difficult because the permitted states are unknown in general.

Therefore, attempts to calculate the partition function were based on
standard volume tessellations, such as the Voronoi and Delaunay
tessellations \cite{Voronoi1908,Delaunay1934,Aste2008,Aste2007}.  A
possible alternative is an arch-based approach \cite{Slobinsky2015}
which a priori takes only stable configurations into account.
Blumenfeld and Edwards proposed a physically motivated tessellation
based on the quadron construction
\cite{Ball2002,Blumenfeld2003,Blumenfeld2006,Blumenfeld2014}.  In
principle, quadrons are used as quasi-particles describing the
structure of the granulate at any arbitrary position within the system
in a distinct way. It was mentioned in the literature that the ideal
quadron tesselation fails in the presence of non-convex voids in the
granulate \cite{Ciamarra2007}. However, in a system of monodisperse
spheres such non-convex voids vanish by neglecting rattlers
\cite{Blumenfeld2007}.  Even if some non-convex voids remain, they
could be tesselated by convex polygons which repairs the quadron
tessellation, at the price that the number of quadrons increases. As
long as non-convex voids are the exception rather than the rule, they
will not produce significant changes to the calculations.

Under the (very rough) assumption that quadrons occupy
volumes between $V_0-\Delta$ and $V_0+\Delta$ at constant density of states and
that there are no interactions between the quadrons, the partition
function can be calculated explicitly for two dimensional systems (see
\cite{Blumenfeld2003}):
\begin{equation}
  Z= \left( \frac{\sinh(\beta \Delta)}{\beta \Delta} e^{-\beta V_0} 
  \right)^{N \bar z} \label{eq:iqZ},
\end{equation}
where $N$ is the number of particles and $\bar z$ is the mean
coordination number in the granular system.  This approximation is
called the ideal quadron approximation by analogy with the description
of ideal gases in ordinary statistics.
Note that the partition function \eqref{eq:iqZ} is a special version
of a more general ideal-gas-like approach. If one assumes that
the volume is tesselated by a number $\tilde{N}$ of statistically
independent, non-interacting elementary cells and their volume is 
restricted to an interval between a minimal volume $V_0-\Delta$ and a maximal
volume $V_0+\Delta$, without any additional assumption on the nature of
these elementary cells one ends up with Eq.~\eqref{eq:iqZ}, on replacing
$N \bar{z} \rightarrow \tilde{N}$. Contrary to the very general
ideal-gas-like approach, there are some possibilities to go beyond
the interaction-free situation in the quadron approach, which will be
considered in future work. 

Using \eqref{eq:meanVolume}, the ideal quadron
prediction for the mean volume is obtained:
\begin{equation}
  \mw V =N \bar z \left( V_0+\frac{1}{\beta} - \Delta
    \coth( \beta \Delta) \right)  \label{eq:meanvolumeideal}.
\end{equation}
For the current work, it is helpful to rewrite
\eqref{eq:meanvolumeideal} in terms of the volume fraction. Dividing
\eqref{eq:meanvolumeideal} by the total grain volume $V_g$ results
in
\begin{equation}
  \bar{\phi}^{-1} = \frac{N \bar z V_0}{V_g} + \frac{N \bar z}{V_g \beta}
  - \frac{N \bar z \Delta}{V_g} \coth(\beta \Delta)\>. \label{eq:meanfractionid1}
\end{equation}
In order to specify the free parameters $V_0$ and $\Delta$, we assume
that the limits of $\bar\phi^{-1}$ can be identified with the volume
fractions of random loose packing (rlp) and random close packing (rcp) in the
following way:
\begin{align}
  \phi_{\text{rlp}}^{-1} &= \lim_{\beta \rightarrow 0} \bar\phi^{-1} =
  \frac{N \bar z V_0}{V_g}\>,
 \label{eq:phirlp}\\
 \phi_{\text{rcp}}^{-1}  &= \lim_{\beta \rightarrow \infty} \bar\phi^{-1} 
=  \frac{N \bar z (V_0-\Delta)}{V_g}\>. 
 \label{eq:phircp}
\end{align}
Now we can rewrite \eqref{eq:meanfractionid1} as
\begin{equation}
  \bar\phi^{-1} = \phi_{\text{rlp}}^{-1} + \frac{N \bar z}{V_g \beta}
  - \Delta_\phi^{-1} \coth\left( \beta \frac{V_g \Delta_{\phi}^{-1} }{N \bar z} \right)
 \label{eq:phiidealquadr},
\end{equation}
where
$\Delta_\phi^{-1}=\phi_{\text{rlp}}^{-1}-\phi_{\text{rcp}}^{-1}$.
Below, eq.~\eqref{eq:phiidealquadr} will be useful as a fitting
function for our simulation data.

We remark that the ideal quadron solution is a very rough estimation,
because the main difficulty in calculation of the partition function
\eqref{eq:kanonischeZustandssumme} is filtering out stable
configurations. In the ideal quadron solution, a minimal filtering is
performed in the sense that there is a minimal and a maximal volume
per quadron so that configurations that are either too loose or too
dense are filtered out. However, the number of states between these
limits may be overestimated.  The arch-based approach
\cite{Slobinsky2015} may have the capability to overcome this issue.
However, in the current form no analytical equation such as
\eqref{eq:phiidealquadr} is available. Only numerical solutions under
very simplifying assumptions are on hand, which makes a comparison of
our data with this model difficult.

Note that there is no commonly accepted definition of the states rlp
and rcp. In this paper we use these terms in the sense of
eq.~\eqref{eq:phirlp} and \eqref{eq:phircp}.  An interesting fact is
that in the framework of the ideal quadron model for $\beta
\rightarrow 0$, which is the limit of very high compactivities, the
mean volume per quadron is $V_0$ , not $V_0+\Delta$. States which
would have a mean volume per quadron with $\bar{V}>V_0$ correspond to
a population inversion, i.e., states with negative granular
temperature. It may be speculated that these states correlate with
so-called random very loose packings states \cite{Ciamarra2008}, which
are states only achievable with very special protocols. A verification
of this idea is beyond the scope of the present article.

\section{Simulation method} \label{sec:method}
In our simulation, we extend an existing DEM Code, originally developed
for the investigation of the mechanical properties of granular piles
consisting of two-dimensional polygonal particles \cite{Schinner1999,Schinner2001}.
The dynamics of the $i$th particle's position $\vec{r}_i$ and orientation
$\phi_i$ are described by Newton's and Euler's equations of motion:
\begin{align}   
  m_i \ddot{\vec{r}}_i &= \sum_{j \neq i} \vec{F}_{ij} +
  \vec{F}^{V}_{i} \label{eq:equation_of_motion_tranlation} \>,
  \\
  J_i \ddot{\phi}_i &= \sum_{j \neq i} {M}_{ij} \>.
  \label{eq:equation_of_motion_rotation}
\end{align}
Herein, $\vec F_{ij}$ is the contact force between particle $i$ and
particle $j$, $M_{ij}$ is the corresponding torque acting on the
particle (referred to its center of mass), and $\vec F^{V}_i$ is the
external force acting on the particle (e.g.,~gravity). The mass of
particle $i$ is denoted by $m_i$ and its moment of inertia by $J_i$
\cite{Roul2011,Roul2011a,Roul2011b}. We assume external
torques to be absent.

For fast determination of potential particle contacts, the particles
are surrounded by bounding boxes and we employ an incremental
sort-and-update algorithm \cite{Schinner1999,Schinner2001} to identify
overlapping bounding boxes efficiently.  Whenever the bounding boxes
of two particles overlap, we use a closest-feature algorithm
\cite{Schinner1999,Schinner2001} from virtual reality and robotics
applications \cite{Lin1993} to calculate the polygon distance. In the
worst-case scenario, the computational complexity is $\mathcal{O}(n
\log n)$ where $n$ is the number of polygon features (corners and
edges). However, the typical behavior, whenever a good guess from the
last timestep is available, is the calculation of the polygon distance
in constant time $\mathcal{O}(1)$, independently of the polygon edge
number.

\begin{figure}
 \includegraphics[width=0.73\columnwidth]{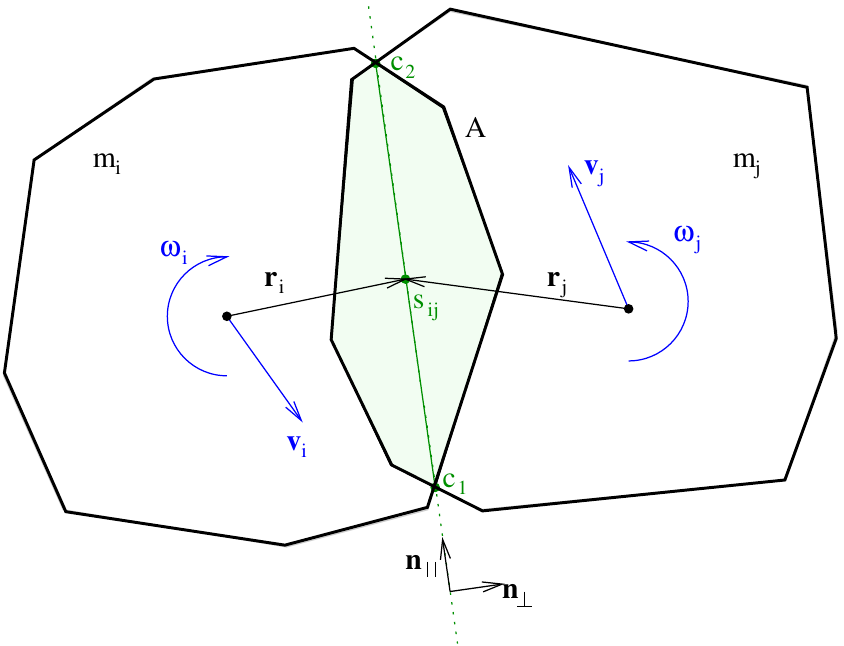}
 \caption{\label{fig:forcegeometrie}(Color online) Sketch of two polygonal particles
   in contact. The normal and tangential direction of a collision is
   determined by the contact line (green), the contact point is
   defined as the middle of the contact line.}
\end{figure}

Figure \ref{fig:forcegeometrie} shows a sketch of two particles in
contact. The contact point $s_{ij}$ between two particles is defined
as the midpoint of the line between $c_1$ and $c_2$.  For the force
calculation, we define some quantities first: the characteristic length
\begin{equation}
 l=\frac{r_i r_j}{r_i+r_j} 
\end{equation}
(note that $l$ is not the length of the contact line), the reduced mass
\begin{equation}
 m_\perp= \frac{m_i m_j}{m_i+m_j}\>,
\end{equation}
and the reduced tangential mass, including the moments of inertia
\begin{equation}
 m_\parallel=\frac{1}{\frac{1}{m_i}+\frac{1}{m_j}+\frac{r_i^2}{J_i}+\frac{r_j^2}{J_j}}
\end{equation}
The relative tangential velocity  at the contact point $s_{ij}$ is 
\begin{equation}
  {v}_\parallel = (\vec{v}_i - \vec{v}_j + (\vec{r}_i \times 
\vec{\omega}_i) - (\vec{r}_j \times \vec{\omega}_j)\ ) \cdot \vec{n}_\parallel.
\end{equation}
where $\vec{v}_{i},\vec{v}_j$ are the velocities of the particles and
$\vec{\omega}_i,\vec{\omega}_j$ are their angular velocities.
Furthermore, we define the effective penetration length as
\begin{equation}
 h_{\text{eff}} = \frac{A}{l}\>.
\end{equation}
Note that $l$ does not change significantly during a collision, so
$h_{\text{eff}}$ is essentially proportional to the overlap area $A$.  The
normal component of the contact force is determined by the equation
\begin{equation}
 F_\perp = \max\left[E h_{\text{eff}} - \gamma \sqrt{E m_\perp} \dot{h}_{\text{eff}}~;~0\right]. \label{eq:normalforce}
\end{equation}
Here $E$ is the two-dimensional Young's modulus and $\gamma$ is the dissipation strength.

The maximum function in \eqref{eq:normalforce} ensures that the normal
force cannot become attractive. Physically, situations where the
dissipative term overcomes the repulsive force correpond to the case
that the particles' separation velocity is faster than the relaxation
velocity of the grain deformation, i.e., the contact between the
particles is lost, before the overlap of the non-deformable model
particles becomes zero \cite{Schwager2007}. Consequently, the contact
force vanishes from this moment on.

We used the Cundall-Strack model \cite{Cundall1979} for modelling
tangential forces. At the beginning of the collision, the
Cundall-Strack force $F^*$ is zero. If this force is given at the
previous timestep, corresponding to time $t_c-\Delta t$, the
Cundall-Strack force at the current timestep at time $t_c$ is
determined by
\begin{align}
   & F^*_\parallel(t_c) = \nonumber \\ 
   & \min \left[ F^*_\parallel(t_c-\Delta t) +v_\parallel(t_c-\Delta t) \Delta t \frac{2}{7} E
    ~;~ \mu F_\perp(t_c)  \right] \>. \label{eq:forcecsspring}
\end{align}
One may visualize this model as a spring between the contact points
being established when a new contact appears and the length of the
spring being limited by the value of Coloumb friction (Coulomb
condition).  If this value is reached, the points of attachment of the
spring start to slide, so that the spring length remains constant. In
this way, the Cundal-Strack model mimics sticking and sliding
friction.  In order to avoid unrealistic oscillations, a damping term
proportional to the velocity is added to the Cundall-Strack force,
which also satisfies the Coloumb conditions. The complete
trangential force  then is
\begin{equation}
  F_\parallel (t_c) = \min \left[ F^*_\parallel(t_c) +  
    v_\parallel \sqrt{\frac{2}{7} E m_\parallel}  ~;~ \mu F_\perp(t_c) \right].
  \label{eq:cundalstrack}
\end{equation}

With \eqref{eq:normalforce} and \eqref{eq:cundalstrack}, all the
contact forces and the resulting torques between the particles
and between particles and walls (which are treated as particles with
fixed position and orientation) are defined. The equations of motion
\eqref{eq:equation_of_motion_tranlation} and 
\eqref{eq:equation_of_motion_rotation} are then solved numerically
using a sixth-order Gear predictor-corrector method \cite{Gear1967}.


\section{Excitation protocols and parameters} \label{sec:protocol}

Inspired by former experimental studies
\cite{Schroeter2005,Zhao2012,Zhao2014}, we implemented two protocols
for exciting the granular matter periodically.  In both protocols, an
excitation period in which the phase space is explored alternates
with a relaxation period in which the grains come to rest completely.
Both protocols are characterized by a control parameter, which we call
tapping parameter.

In all simulations, we used a bidisperse mixture of 1184 regular
decagons, composed of 544 particles with radius $r_1=9\, \text{mm}$
and 640 particles with radius $r_2=6.36\, \text{mm}$. 
Young's modulus was set to $E=1000\, \text{N}/\text{m}$ and
the simulation time step was chosen as $\Delta t=5 \cdot 10^{-5}
\,\text{s}$. The Coulomb friction coefficent was taken to be 
$\mu=0.6$ and for the normal friction coefficent $\gamma=0.5$
was used. The particles mass density was $\rho=0.001\, \text{g}/\text{mm}^2$
and was the same for big and small particles.

With both protocols, the width of the simulation area was $280\,
\text{mm}$.  In the case of the negative $g$ protocol, there is no
lid, for the rotating $g$ protocol the height of the box is $900\,
\text{mm}$. Walls are realised as fixed rectangular particles with the
same Young's modulus and friction coefficients as the mobile particles.

Tejada et.~al.~\cite{tejada2014} pointed out that the size of the
time step in DEM simulations may influence the width and the shape
(but not the mean) of the determined probability distributions, even if
the time step appears sufficiently small using common criteria.  In
order to make sure that this effect does not influence our results, we
repeated some of the simulations with a time step of $\Delta t=
10^{-5} \,\text{s}$. We found that the mean volume fraction, the volume
fraction variance and also the shape of the volume fraction
distribution did not change on reduction of the time step. 

\begin{figure}
 \includegraphics[width=\columnwidth]{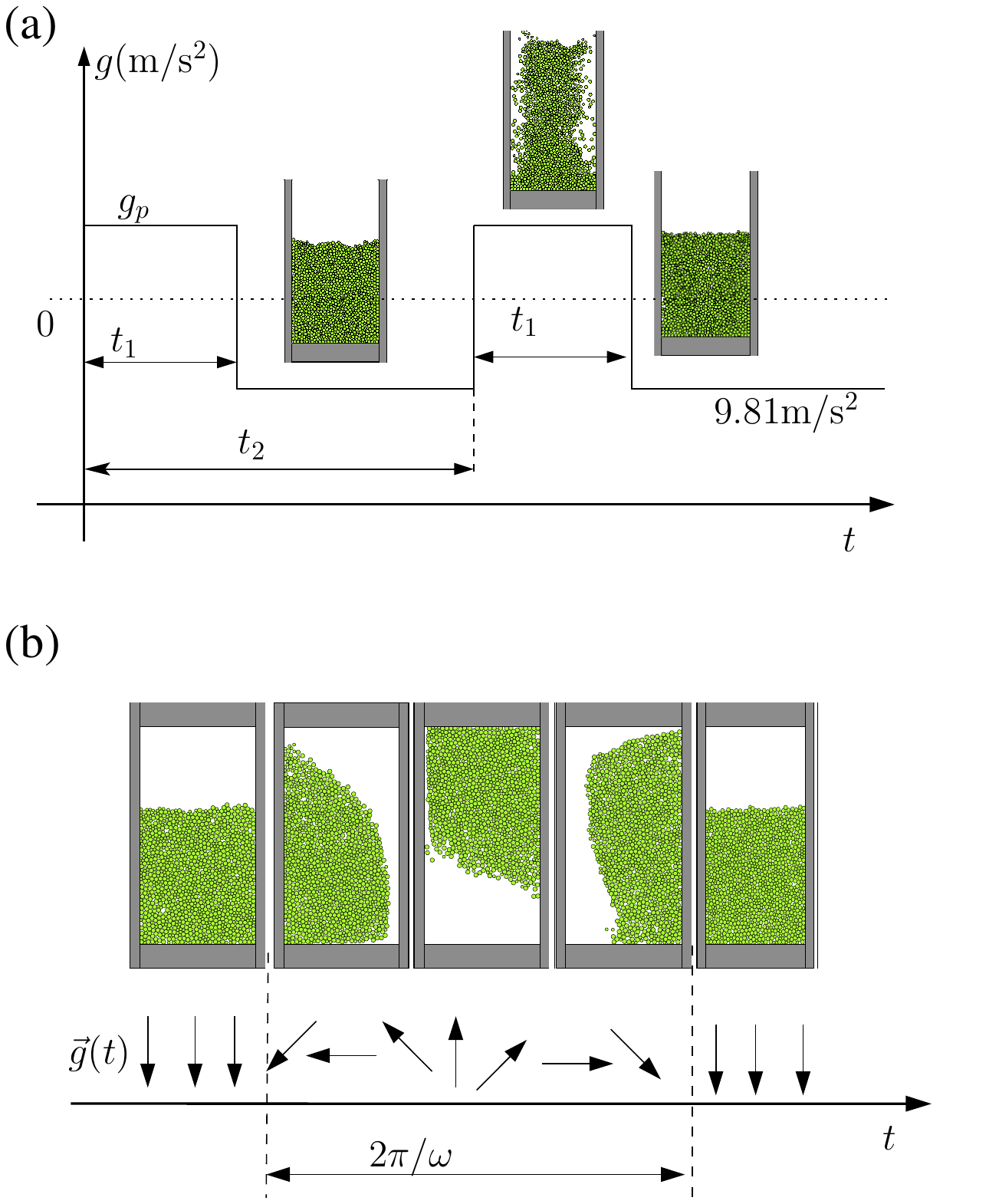}
 \caption{\label{fig:protocol}(Color online) a) The ``negative g'' protocol: For a
   short time interval $t_1$ the gravity is turned upward with a
   prescribed magnitude. Afterwards, the granular system has time to
   come  to rest completely, before the next pulse starts.  b) The
   ``rotating g'' protocol: The gravitational acceleration vector
   performs a complete rotation. Afterwards, the granular system comes
    to rest completely before the next round starts.}
\end{figure}

The first protocol, called ``pulses of negative gravitation'' or
``negative g'' protocol, is illustrated in figure \ref{fig:protocol}a.
Most of the time, the granulate is at rest in a container under
normal gravitation, but for short time intervals, the direction of
gravitational acceleration is reversed. The time dependence of the
gravitational acceleration $g(t)$ is taken to be
\begin{equation}
  \vec{g}(t)= 9.81~\text{m}/\text{s}^2 \vec{e}_y\cdot\begin{cases} ~g_p & \tilde{t} < t_1 
    \\
    -1 &  \text{otherwise} \end{cases}\>.
\end{equation}
Here $\tilde{t}=t \bmod T$ is the simulation time $t$ modulus the
period $T$ of the protocol and $t_1$ is the duration of a pulse.  We
took the pulse length to be $t_1=0.1\,$s and the period $T=3\,$s.  The
relaxation period was chosen so that for the biggest excitation
amplitudes the resulting volume fraction variations due to remaining
kinetic energy in the system are smaller than $10^{-5}$ which is 100
times smaller than the typical magnitude of the observed static volume
fraction fluctuations.  The pulse amplitude $g_p$ was used as a
tapping parameter. Due to Einstein's equivalence principle, this
protocol is equivalent, per period, to
a 
downward acceleration $(g_p+1)g$ for an interval $t_1$ and a
subsequent stoppage in the gravitational field for a time span
$T-t_1$, taken long enough for the granulate to come to rest. Here and
in the following when we write $g$, we mean $g=9.81\, \text{m/s}^2$.

Figure \ref{fig:protocol}b illustrates the second protocol. The
granulate is at rest in a closed box.  Then the gravitational
acceleration performs a complete rotation followed by a relaxation
period.
\begin{equation}
\vec{g}(t)=9.81 \text{m}/\text{s}^2  \cdot\begin{cases} \sin(\omega t) \vec{e}_x - \cos(\omega t) \vec{e}_y & \tilde{t} < \frac{\omega}{2 \pi} \\ -\vec{e_y} & \text{otherwise} \end{cases} \>. 
\end{equation}
As in the negative g case, $\tilde{t}=t \bmod T$ is the simulation
time $t$ modulus the period $T$. The angular frequency $\omega$ is the
control parameter of the protocol. Of course $T>2 \pi/\omega$ must be
fulfilled. In our simulations, we used $T=2\pi/\omega+3\,$s, with
$\omega\gtrsim 4 \,\text{s}^{-1}$.  For both protocols, we tested
whether or not segregation of the particles occurs by measuring the
cumulative number of small and big grains as function of the column
height $n(h)$ (i.e.~the number of small or big particles with $z$
coordinates smaller than $h$).  These curves are straight lines and
do not change during the simulations, except for fluctuations.
Therefore we conclude that in our simulations segregation did not take place.


\section{Determination of compactivity, fluctuations and the partition
  function} \label{sec:anwendung} For both protocols and each choice
of the tapping parameter the simulations ran for $2500-3000$
excitation and relaxation periods. Note that this relatively high
number of taps is necessary to draw serious conclusions for the
systems used here.  While the confidence interval of the mean volume
fraction became sufficiently small after some hundred taps, this was
not the case for the estimated variance. In test simulations, where
only a few hundred taps were considered, the uncertainty of the
estimated variance \cite[pp. 771--772]{bronstein1999} was as big as
the domain of the measured variance itself.

\begin{figure}[h]
 \includegraphics[width=0.88\columnwidth]{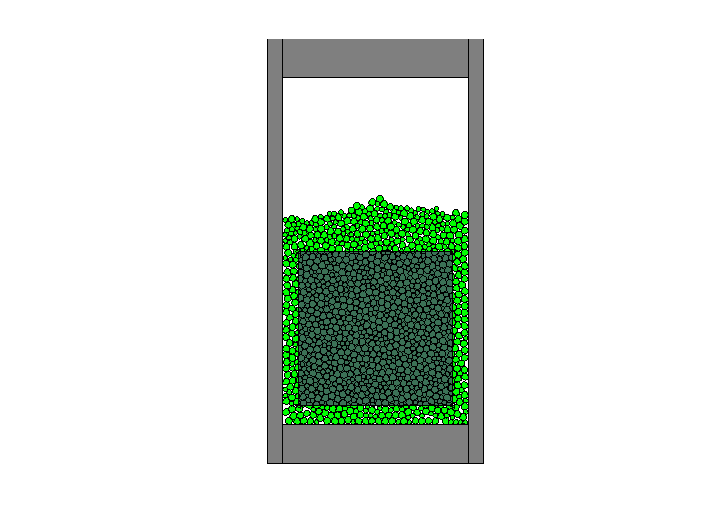}
 \caption{\label{fig:testvolume}(Color online) A typical situation when the
   granulate is at rest. The region shaded grey is the test volume,
   used for volume fraction determination.}
\end{figure}

Immediately before the relaxation period ended, we determined the
volume fraction of the granulate by measuring the fraction of solid
particles in a test volume. The test volume shown in figure
\ref{fig:testvolume} is a square with edge length of $400\, \text{mm}$
and was chosen in such a way that some layers of particles were
between the borders of the test region and the walls.

It was pointed out \cite{Gago2015} that test volumes must be big
enough to avoid size depended effects.  To make sure that our test
volume is sufficiently large, we divided it into two neighbouring
columns. The relative difference of mean volume fraction (after
2500 taps) between the columns and the entire test volume was always
smaller than $0.01\,$\% and the deviation of the volume fraction fluctuation
ratio from $\sqrt{2}$ was always lower than $1\,\%$.

\begin{figure}
 \includegraphics[width=\columnwidth]{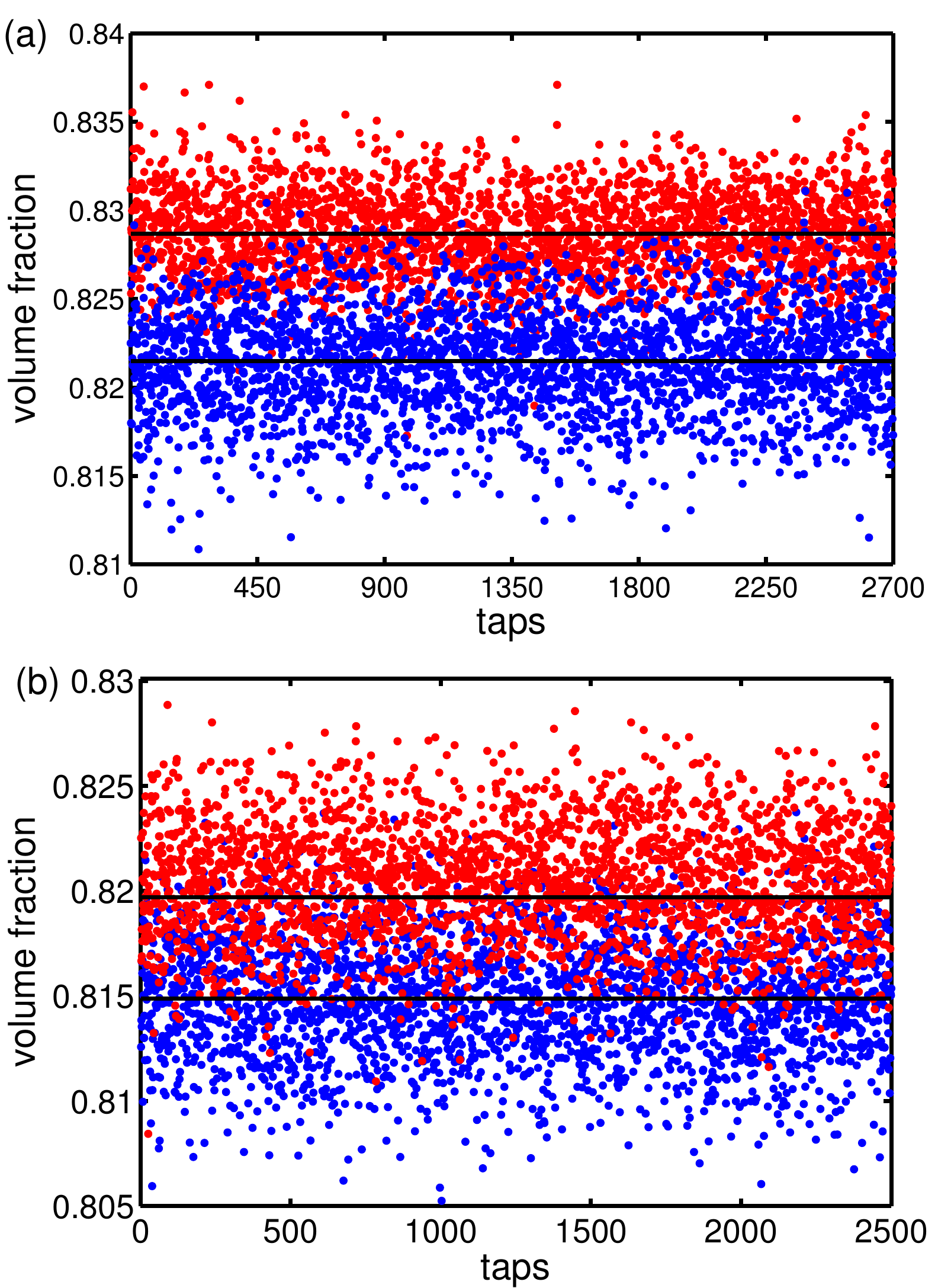}
 \caption{\label{fig:timeseries}(Color online) Exemplary time series for both
   protocols. The points are the volume fractions calculated from
   simulation data and the solid lines are the corresponding mean
   values.  The tapping parameter is for the ``negative $g$'' protocol
   (a): $g_p g=6 \text{m/s}^2$ (red points) and $g_p g=28 \text{m/s}^2$
   (blue points) and for the ``rotating $g$'' protocol $\omega=0.75 \cdot
   2\pi~\mathrm{s}^{-1}$ (red points) and $\omega=2 \cdot
   2\pi~\mathrm{s}^{-1}$ (blue points).}
\end{figure}

Figure \ref{fig:timeseries} shows some typical time series for both
protocols. The mean volume fraction reaches a steady state value very
quickly (after approximately $<10$ taps) and then only fluctuations
around its mean value occur.

\begin{figure}
\includegraphics[width=\columnwidth]{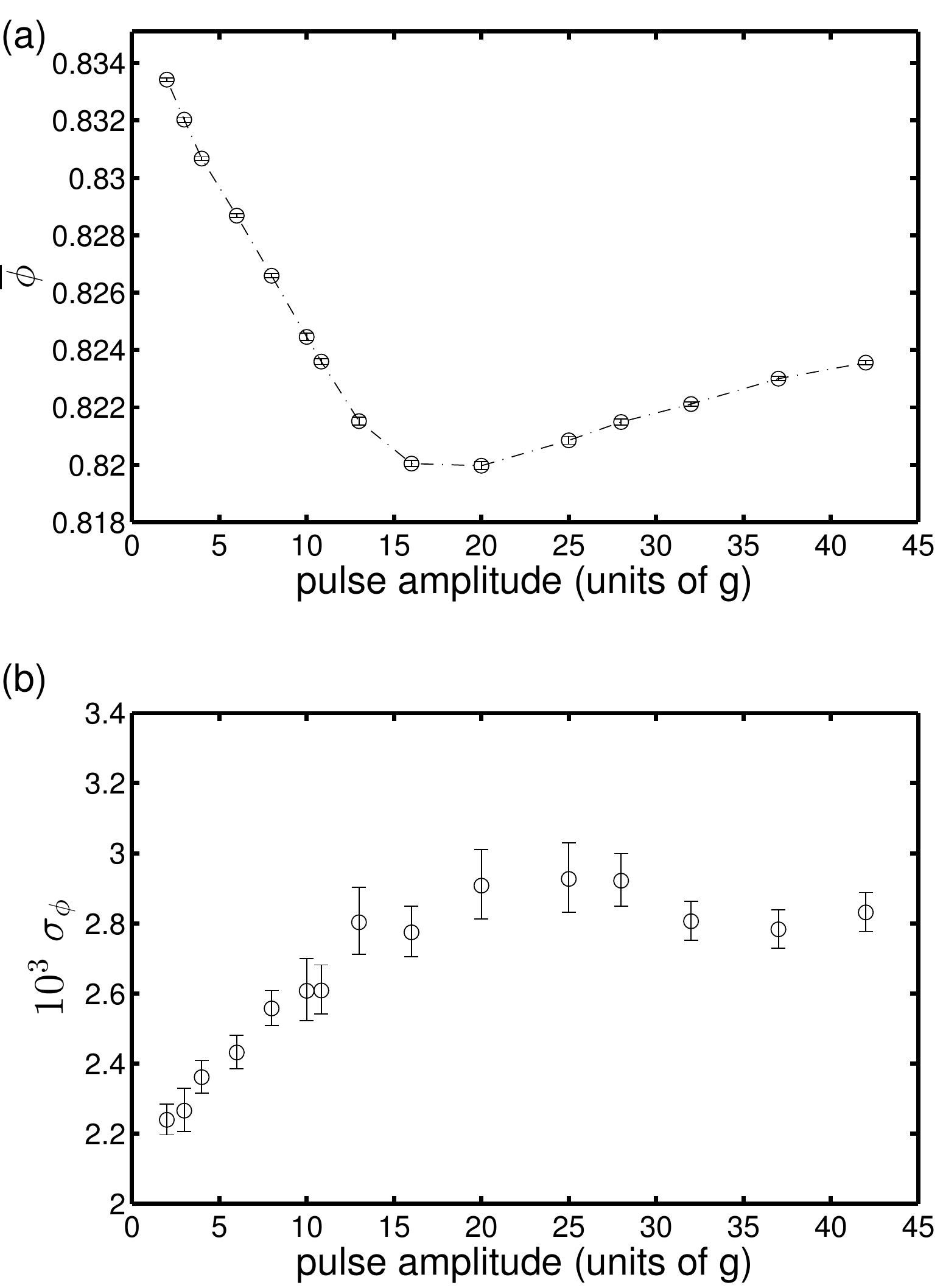}
\caption{\label{fig:volfracandfluc_g}Simulation results for the
  ``negative g'' protocol: (a) the mean volume fraction as function of
  the pulse strength in units of earth acceleration is depicted, (b) shows the standard deviation of
  the volume fraction fluctuations. The error bars correspond to a
  confidence interval of 95 percent.}
\end{figure}

In Fig.~\ref{fig:volfracandfluc_g}, the mean value of the volume
fraction $\bar{\phi}$ and the standard deviation of the volume
fraction distribution, characterizing the volume fraction fluctuation
strength, are shown. Initially, the volume fraction decreases with
increasing pulse strength until it reaches a local minimum around
$g_p=20$. Afterwards, the volume fraction increases with the pulse
strength. Similar behaviour for tapped granulates at high tapping
strengths was described in \cite{Gago2009,Pugnaloni2010}.  A possible
explanation is the interplay between two competing effects. First, the
stronger the pulse of negative gravitation, the more the granulate is
whirled around, i.e.~the looser the resulting packing should get. On
the other hand, the stronger the pulse, the higher the granulate
flies, therefore the higher its impact velocity when it reaches the
bottom, resulting in stronger compaction during the relaxation period.
The first effect dominates for relatively small values of $g_p$, the
second effect becomes more important for stronger pulses.  At the same
value of $g_p$, where the volume fraction is minimal, the volume
fraction \emph{fluctuations} have a local maximum (see
Fig.~\ref{fig:volfracandfluc_g}b). Similar results were found in work
by Pugnaloni \emph{et al.}~\cite{Pugnaloni2010}. 

\begin{figure}
 \includegraphics[width=\columnwidth]{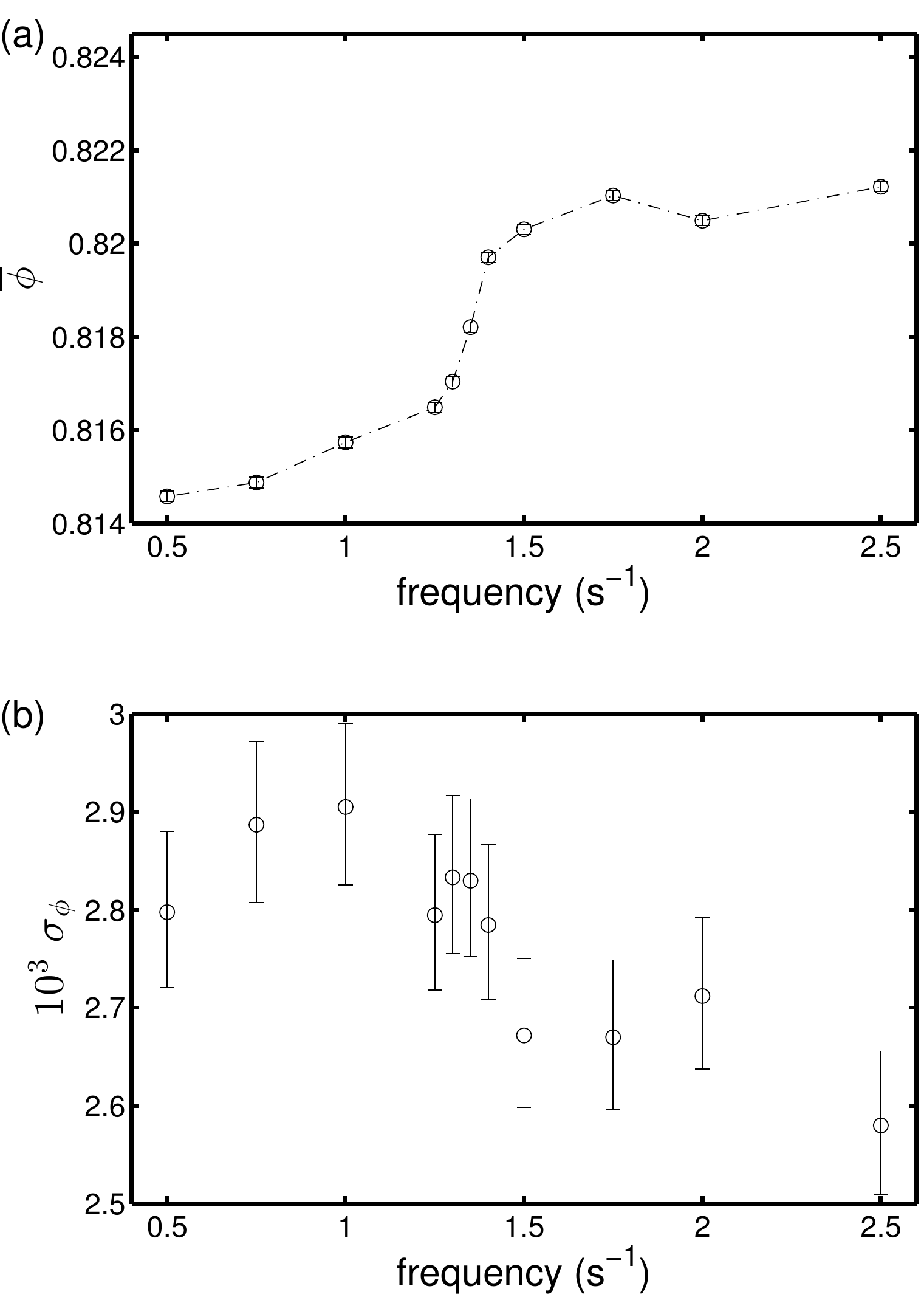}
 \caption{\label{fig:volfracandfluc_rot} Simulation results for the
   ``rotating g'' protocol: Part (a) shows the mean volume fraction as
   function of rotation frequency $f=\omega/(2\pi)$ and (b) shows the standard
   derivation of the volume fraction fluctuations. The errorbars
   correspond to a confidence interval of $95$ percent.}
\end{figure}

The corresponding results for the ``rotating g'' protocol are shown in
figure \ref{fig:volfracandfluc_rot}. With increasing frequency the
mean volume fraction shows a crossover from $\bar{\phi}\simeq0.814$ to
$\bar \phi \simeq 0.821$.  When the frequency is very small, a
complete rotation takes a long time and the granulate behaves in a
quasistatic way. It is clear that in this case a further decrease of
the frequency will not have a significant influence on the resulting
volume fraction. Therefore, for small frequencys the $\phi(\omega)$
curve should approach a plateau. When the frequency increases, the
rotation gets faster, the granulate is more strongly whirled around
and becomes looser. When the frequency increases further, the time per
rotation gets smaller which compensates the increase of swirling due
to faster rotation. For very high frequency, one expects that the
system reaches an irreversible regime, because it does not have time
to respond to the rotation pulse and will largely stay in the initial
configuration - but this regime is not reached in our simulations.

In order to determine the granular temperature of the samples with the
overlapping histogram method, the probability density distribution of
the volume fraction must be estimated. Notwithstanding the name of the
approach, we used the more sophisticated kernel density estimation
method \cite{Rosenblatt1956,Parzen1962} instead of histograms in order
to obtain an approximation for the probability density. A normal
kernel was employed and the bandwidth was chosen according to
Silverman's rule of thumb \cite{Silverman1998}. In appendix
\ref{append:kde}, a short description of the approach is offered.

\begin{figure}
 \includegraphics[width=\columnwidth]{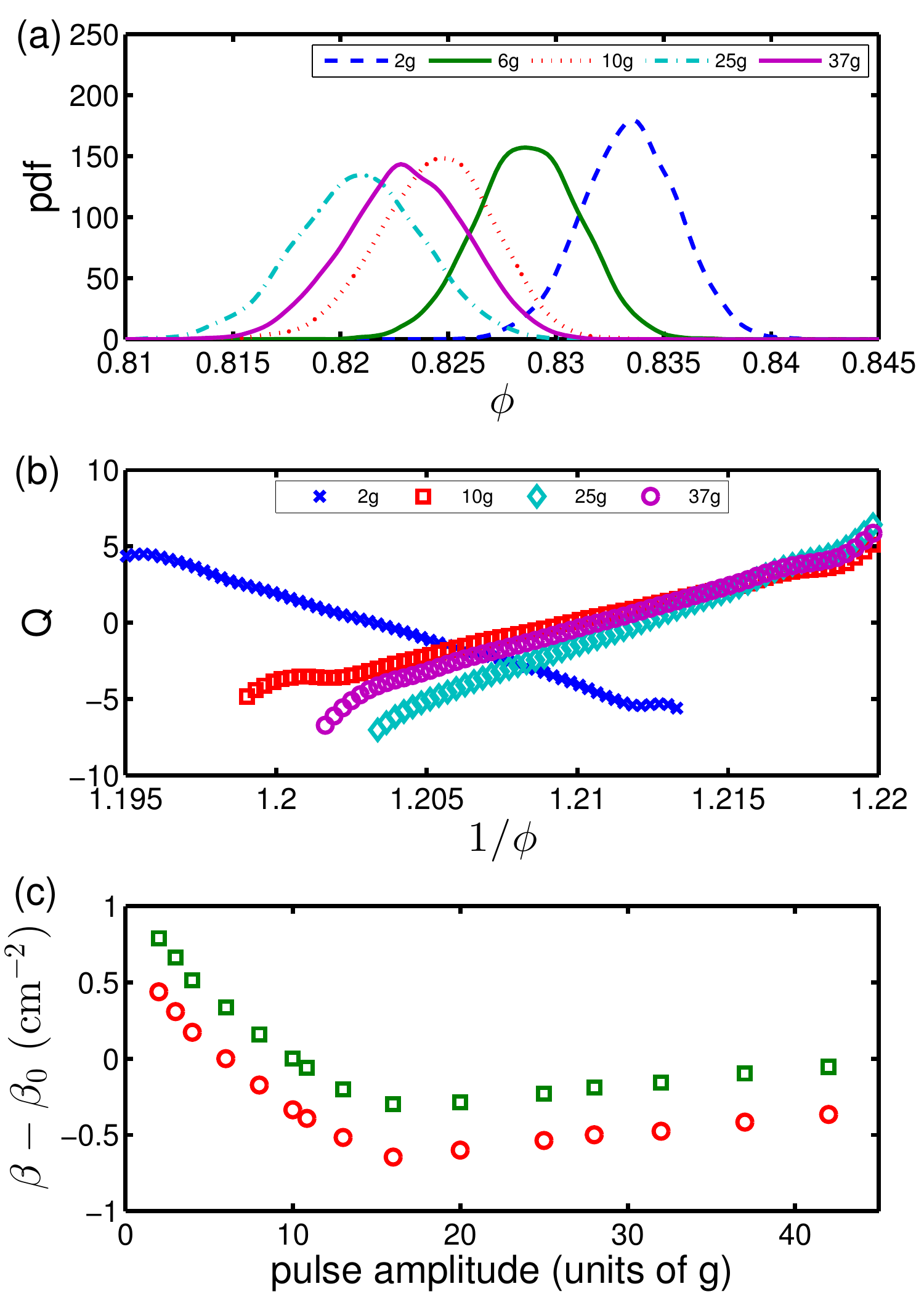}
 \caption{\label{fig:oh_neg_g}(Color online) (a) Probability density function of the
   volume fraction for the ``negative g'' protocol, estimated with the
   kernel density method, for different values of the pulse strength.
   (b) The quantity $Q$ from equation \eqref{eq:overlappinghistofit}
   for some values of the tapping parameter, where the distribution
   with $g_p=6$ was chosen as denominator in \eqref{eq:Qdefinition}.
   (c) The compactivity calculated with the help of the overlapping
   histogram method using the distribution with $g_p=6$ (green, squares) and
   $g_p=10$ (red, circles) as denominator in \eqref{eq:Qdefinition}. }
\end{figure}

\begin{figure}
 \includegraphics[width=\columnwidth]{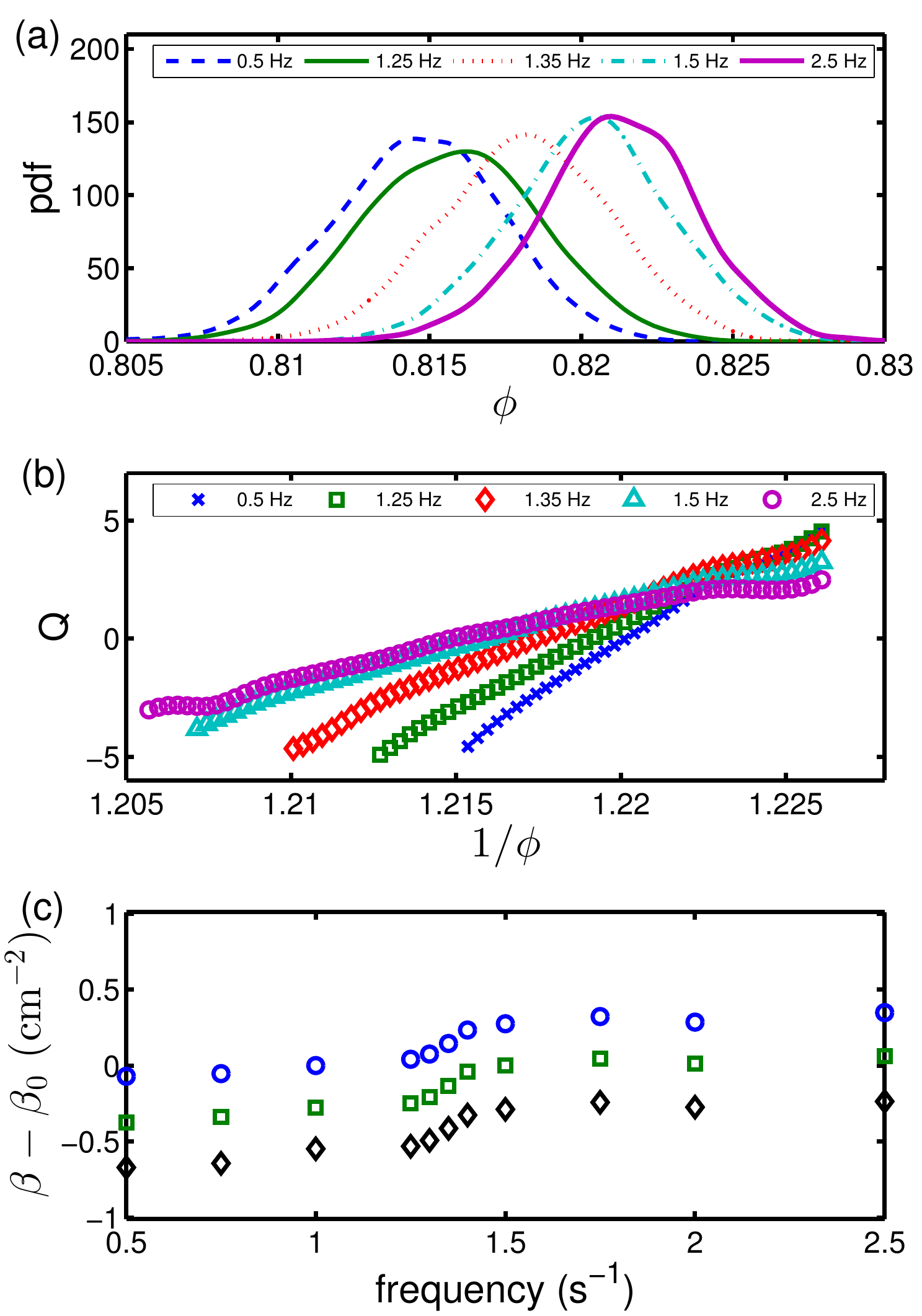}
 \caption{\label{fig:oh_rot}(Color online) (a) Probability density function of the
   volume fraction for samples of the ``rotating g'' protocol,
   estimated with the kernel density method, for different values of
   the rotation frequency. (b) The quantity $Q$ from equation
   \eqref{eq:overlappinghistofit} for some values of the tapping
   parameter using the distribution with $g_p=6$ of the ``negative g''
   protocol as denominator in \eqref{eq:Qdefinition}.  (c) The
   compactivity calculated with the help of the overlapping histogram
   method using the distribution from the ``rotating g`` protocol with
   $f=0.5 Hz$ (blue, circles), $f=1.25 Hz$ (green, squares) and the
   distribution from the ``negative g'' protocol with $g_p=6$ (black,
   diamonds) as denominator.}
\end{figure}

Some of the determined probability density distributions are shown in
Fig.~\ref{fig:oh_neg_g}a for the ``negative g'' protocol and in
Fig.~\ref{fig:oh_rot}a for the ``rotating $g$'' protocol. In order to
test if the distributions are Gaussian, we used a chi-square test
\cite{bronstein1999} with a significance level of $5$\%. The null
hypothesis that the data is normally distributed was rejected for all
samples \footnote{Note that in early test simulations with approx. 100
  taps, the null hypothesis was not rejected, one needs enough data to
  decide about this hypothesis.}. Although Gaussians may still be good
approximations to the central region of our distributions, we take
this as evidence for a statistical mechanical origin of these
distributions rather than their emergence from some unknown additive
process not related to phase space exploration. Therefore, we believe
that our data do not give spurious results due to the pitfall
described in appendix \ref{app:gauss}. 

We also show in appendix  \ref{app:gauss}
that while the presence of Gaussian distributions may lead to false
positive results regarding the validity of Edwards' theory
\cite{McNamara2009}, this is by no means true for all Gaussian
distributions: in the limit of large numbers, the canonical
distribution corresponding to a fixed compactivity must become
Gaussian as well, and this distribution obviously satisfies Edwards'
theory by definition. We take the fact that the tails of our
distributions are non-Gaussian together with the verification of
\eqref{eq:overlappinghistofit} as an indication that the theory works.
This interpretation is supported by other studies of very similar
systems where the volume-per-particle distribution (which is not a sum
and therefore not subject to the central limit theorem) was found not
to be Gaussian \cite{Zhao2014,McNamara2009} even in the central
region.
  
In terms of volume fraction, equation \eqref{eq:overlappinghistofit}
reads:
\begin{align}
  Q &=\ln \frac{P(\phi,\beta_1)}{P(\phi,\beta_0)}
  \nonumber \\
  &=-(\beta_1-\beta_0) \frac{V_g}{\phi}-\ln
  \frac{Z(\beta_1)}{Z(\beta_0)} \label{eq:Qfitfrac}.
\end{align}
We note that \eqref{eq:Qfitfrac} is interpreted in terms of a
canonical ensemble, which implies the number of particles to be a
constant. This means that the total volume of the grains $V_g$ in the
test volume must be a constant \footnote{Equation
  \eqref{eq:overlappinghistofit} could be interpreted as describing a
  grand canonical ensemble as well and would be evaluable for variable
  particle number. With \eqref{eq:Qfitfrac} this is problematic,
  because then the compactivity would be multiplied by two factors
  $V_g$ and $\phi$ that might both vary.}.  In principle, one would
have to adapt the test volume size $V_T$ used in determining the
volume fraction distribution to the measured mean volume fraction
$\bar \phi$ in such a way that $V_g=V_T \,\bar \phi$ remains constant.
The size of the test volume would then be a function of the mean
volume fraction.  This could become important, if very small test
volumes were used. The standard deviation of the volume fraction
distribution should decrease with increasing test volume as $\sigma
\propto V^{-1/2}$, which follows from \eqref{eq:volflukbetarelation},
while $\bar \phi$ does not depend on the system size. By using a
constant test volume $V_{T}$, the magnitude of differences in $V_g$
due to different volume fractions is bounded by $\Delta V_g =
V_T(1/\phi_\text{rcp}-1/\phi_\text{rlp})\simeq0.05 V_T$. If the test
volume is large compared to the size of the particles, the relative
error which is made by using a constant test volume, assuming that the
cumulative grain volume is constant (in spite of the different volume
fractions), corresponds to approximately $(2-3)\%$. This is smaller
than the confidence interval of $\sigma_\phi$ due to the finite sample
size and therefore negligible.

By choosing a reference probability distribution which is used as the
denominator in \eqref{eq:Qfitfrac} we are able to determine the
inverse compactivity and the logarithm of the partition function up to
additive constants, which are the unknown inverse compactivity of the
reference distribution and the unknown logarithm of the partition
function thereof, respectively. In order to avoid errors due to
insufficient data in the tails of the distribution, we evaluated the
slope of $Q$ only in regions where the value of each distribution
involved in the calculation of \eqref{eq:Qfitfrac} is bigger than $5$\% of
its maximum. 

Figures \ref{fig:oh_neg_g}b and \ref{fig:oh_rot}b show the quantity
$Q$ for some samples as function of the inverse volume fraction, where
the distribution with $g_p=6$ of the ``negative g'' protocol was
chosen as the reference distribution, because it has sufficient
overlap with all the other distributions. We tried several other
distributions as references, too. So long as the distribution
overlap was big enough, we always found a linear relation between $Q$
and the inverse volume fraction $\phi^{-1}$. This may be interpreted
as suggestive of the validity of Edwards' assumptions. Note that the
deviations from the straight line on the left and right ends of the
curves are due to insufficient number of sampling data points in that
region.  Therefore the estimated probability density function behaves
like the tails of the sampling kernel, which is reflected also in $Q$.
This behaviour is a mathematical artefact and not a systematic
deviation from a straight line.

From \eqref{eq:Qfitfrac}, it follows immediately that the values
determined for $\beta$ using different samples as reference
distribution may only differ by an additive
constant. 
This prediction holds for the ``negative g'' protocol as is
demonstrated in Figure \ref{fig:oh_neg_g}c. There, the samples with
$g_p=6$ and $g_p=10$ were used as reference distributions. The same is
true for the ``rotating g'' protocol (Fig.~\ref{fig:oh_rot}c) when we
used samples corresponding to different values of the rotating
frequency. It even applies if we use samples from the ``negative g''
protocol as reference distribution, in agreement with the fact that
the predictions of Edwards' theory should be protocol independent.

\begin{figure}
 \includegraphics[width=\columnwidth]{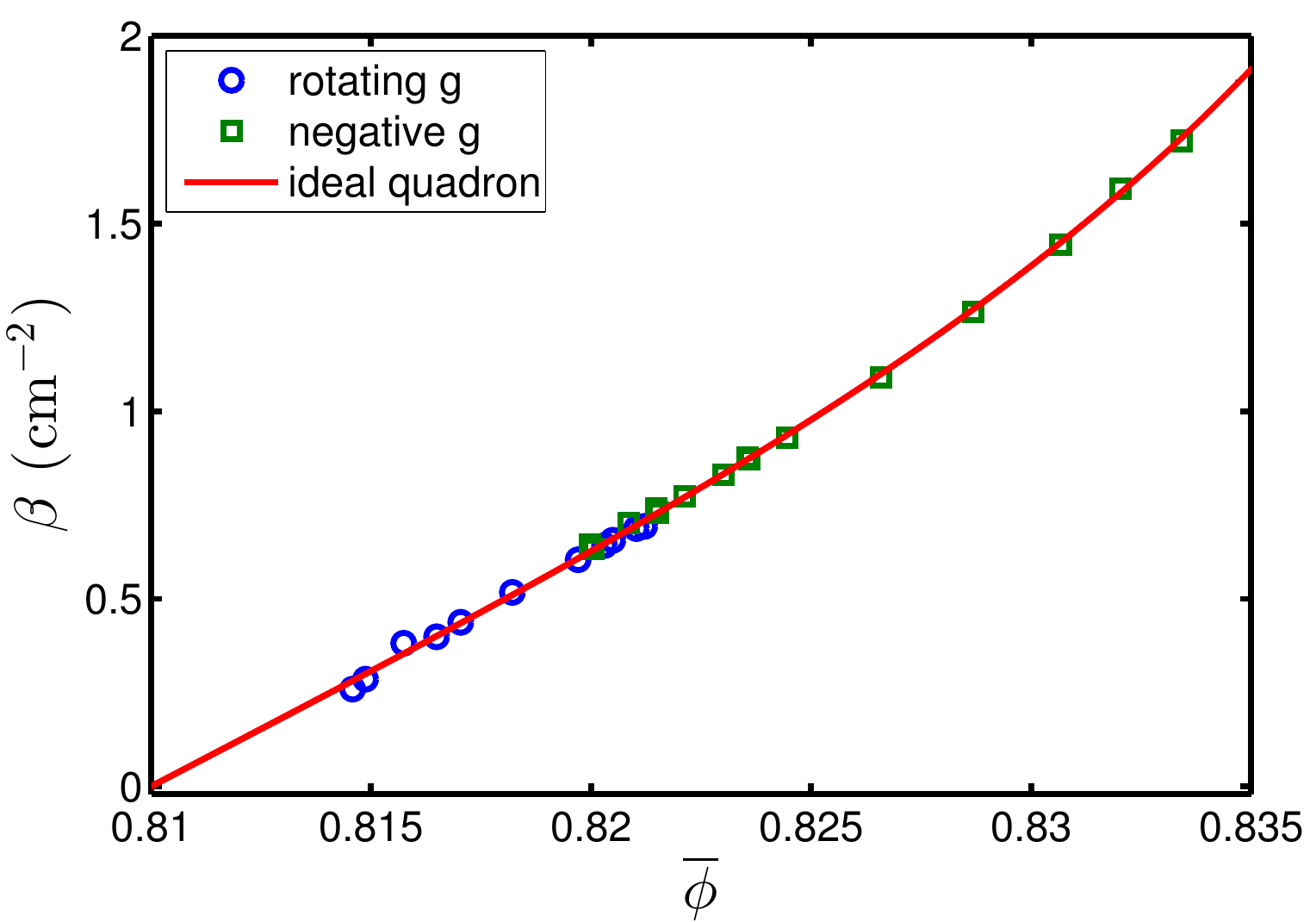}
 \caption{\label{fig:betaoverphi}(Color online) The compactivity as function of the
   volume fraction density for the ``negative g'' protocol (square)
   and for the ``rotating g'' protocol(diamonds). The values were
   corrected with the additive constant $\beta_0$ determined from the
   ideal quadron fit.  The solid line is a fit to the ideal quadron
   solution \eqref{eq:phiidealquadr}}. 
\end{figure}

From now on, we use the distribution from the ``negative g'' protocol
with $g_p=6$ as reference distribution for all the following calculations.
Figure \ref{fig:betaoverphi} shows the values  determined  for the
inverse compactivity for both protocols as function of the volume
fraction. All the data, whether obtained from the branch left of the minimum or the
branch right of the minimum in the $\phi(g_p)$ curve of the ``negative g'' protocol
(cf.~Fig.~\ref{fig:volfracandfluc_g}) or else from the ``rotating g''
protocol is fitted by the same function $\beta(\phi)$. This is a
strong indicator for the applicability of Edwards' theory to our
samples.

In order to fit our results to an analytical function, we used the
ideal quadron solution \eqref{eq:phiidealquadr}. The values of random
loose packing (rlp) and random close packing (rcp) were assumed to be
$\phi_{\text{rlp}}=0.81$ and $\phi_{\text{rcp}}=0.855$. The choice of
these values was pragmatic. To our knowledge, there are no studies about
the exact values for random close and random loose packing for
bidisperse decagons.  Furthermore the values of rlp and rcp will
depend on on the size ratio between the particles. Therefore, we used
plausible values obtained for (bidiperse) disks
\cite{Berryman1983,Silbert2010,Zhao2014}. We checked that the
influence of the values of random loose and random close packings on
the values of the obtained parameters are smaller than $20\,\%$, as long
as the values are in the plausible interval.

$N \bar z/V_g$ was used as fitting parameter. Since we can determine
the inverse compactivity only up to an additive constant from the
overlapping histogram method, we made the replacement $\beta
\rightarrow \tilde{\beta}+\beta_0$ in \eqref{eq:phiidealquadr}, where
$\tilde{\beta}$ is the value determined from the simulations and
$\beta_0$ is taken as an additional fit parameter. Note that the
determination of $\beta_0$ is possible as we assumed that the inverse
compactivity is fixed at random loose and random close packing and
that this is not specific to the ideal quadron fit. We emphasize that
the parameter $\beta_0$ only shifts the whole curve shown in
Fig.~\ref{fig:betaoverphi} upward or downward and is the same for both
protocols.  Since the \emph{same} value of $\beta_0$ was added to \emph{every} data
point obtained from the overlapping histogram method for the
comparison of the fit function with the simulation data in
Fig.~\ref{fig:betaoverphi}, this value does not influence the
conclusion that the compactivity is the same for both protocols.
The fit describes the simulation data very well, as is seen in 
Fig.~\ref{fig:betaoverphi}. However, we find as fitting value $N\bar
z/V_g=556.51\, \text{m}^{-2}$, which differs by about a factor of $40$
from the values used in the simulation. This might be understood by
assuming that approximately 40 quadrons constitute a statistically
independent unit in the granular ensemble. This issue certainly needs
further study. We also tried to fit $\phi_{\text{rlp}}$ and
$\phi_{\text{rcp}}$ and using the known particle number in
\eqref{eq:phiidealquadr} but this leads to an unphysical value for
random close packing of about $\phi_{\text{rcp}}\simeq1.2$.

When we used only the data obtained by one of the protocols to
determine the parameters of the ideal quadron solution, the solution
fitted the data of the other protocol.  The relative deviation of the
obtained fitting parameters is smaller the $2\,\%$. Because this
results in curves that are indistinguishable to the eye, only the fit
using the whole dataset is presented in in Fig.~\ref{fig:betaoverphi}.
Nevertheless, this means that we can predict the $\beta(\phi)$ curve,
for the ``rotating $g$'' protocol using the data determined from the
``negative $g$'' protocol and vice versa. However, for all
compactivities determined, the same \emph{reference} distribution was
used so the data of the fit employed for the one and the other
protocol was not entirely independent.

\begin{figure}
 \includegraphics[width=\columnwidth]{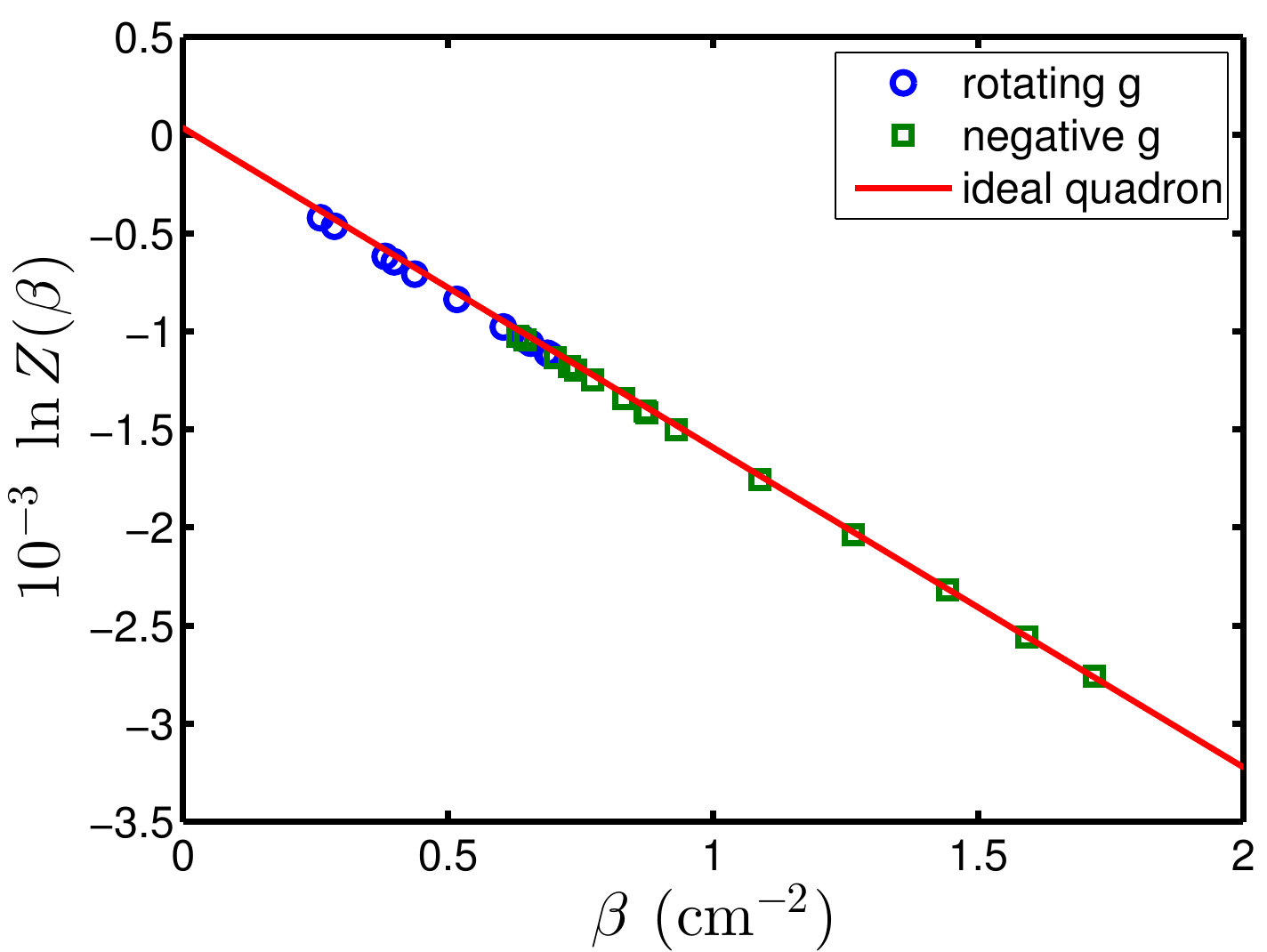}
 \caption{\label{fig:partitionfuntion}(Color online) The logarithm of the partition
   function, determined from the overlapping histogram
   method(symbols). The solid line is the ideal quadron solution.}
\end{figure}

While the slope of $Q$ allows us to determine the inverse
compactivity, the axis intercept $B$ determines the logarithm of the
partition function
\begin{equation}
B=-\ln Z(\beta)+\ln Z(\beta_0). \label{eq:interception}
\end{equation}
If the assumptions leading to \eqref{eq:Qfitfrac} are correct, the
partition function of the ideal quadron solution \eqref{eq:iqZ} should
describe the found intercept without additional fitting. The
parameters $\Delta$, $V_0$, $Nz$ are directly related to the
parameters determined via Eqs.~\eqref{eq:phirlp} and
\eqref{eq:phircp}.  As Fig.~\ref{fig:partitionfuntion} shows, the
numerically determined values and the ideal quadron solution fit very
well, independently of the protocol and of the branch in the
``negative g'' protocol.

\begin{figure}
 \includegraphics[width=\columnwidth]{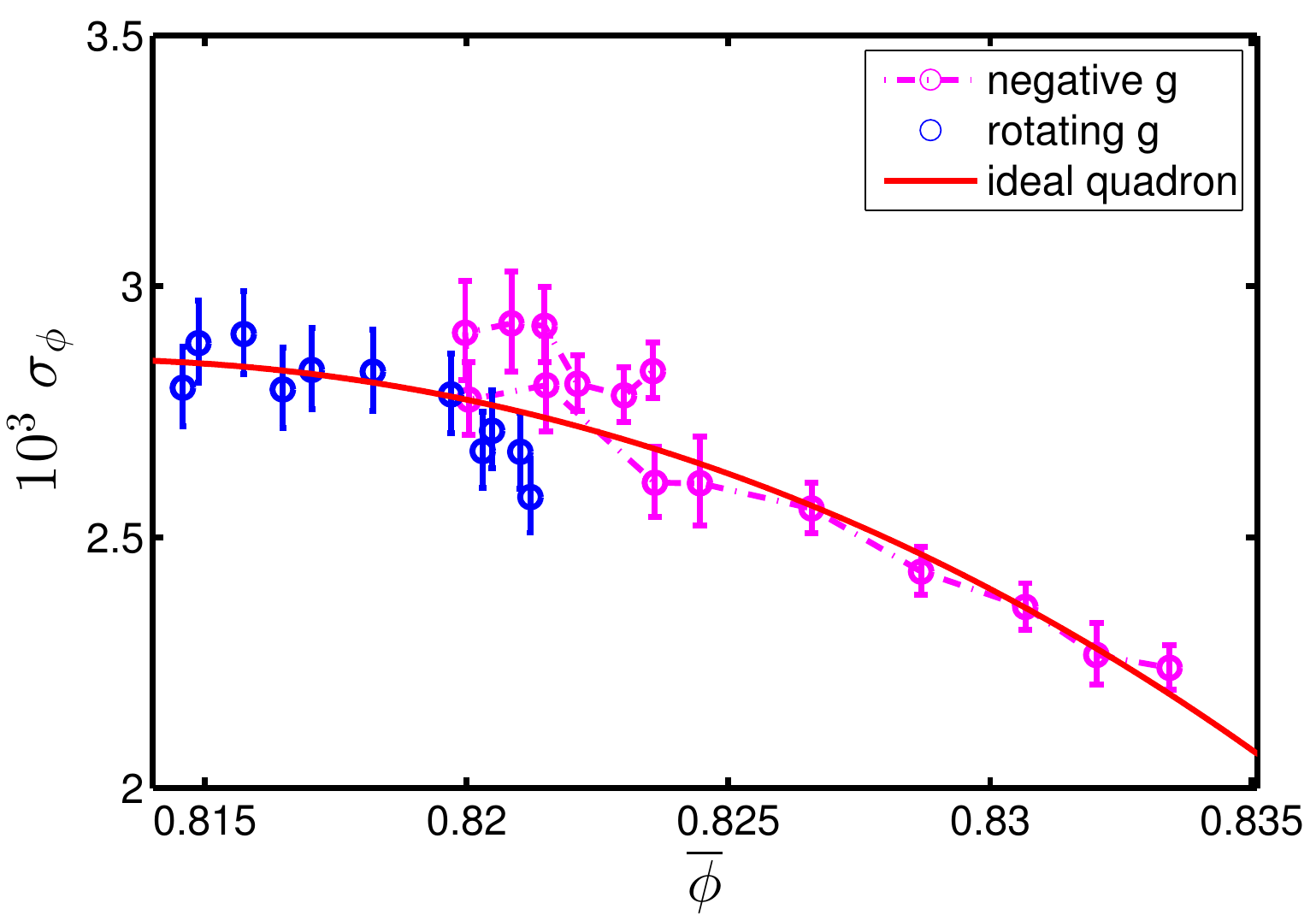}
 \caption{\label{fig:sigma_phi_plot}(Color online) The volume fraction fluctuation
   (standard deviation) as function of the mean volume fraction. The
   blue points correspond to the ``rotating g`` protocol and the
   magenta points and lines corresponds to the negative g'' protocol.
   The solid line is the ideal quadron fit.}
\end{figure}

Using \eqref{eq:phiidealquadr} in \eqref{eq:volflukbetarelation}, we
get the relationship between the mean volume fraction and its
fluctuations. The result together with simulation data obtained directly
is shown in figure \ref{fig:sigma_phi_plot}. The data is in good
agreement with the theory for both protocols.

We mention that in a former study which used vertically tapped
monodisperse regular polygons \cite{Carlevaro2011}, a maximum in the
$\phi-\sigma$ curve was reported which coincided with an inflexion
point in the impulse strength - volume fraction curve.  In our
simulation we do not see this effect, also in experimental work about
bidisperse two dimensional systems such a maximum was not observed
\cite{Zhao2014}.  It might be speculated that crystallization
effects that occur frequently in two-dimensional monodisperse
systems were responsible for the occurrence of the maximum in
Ref.~\cite{Carlevaro2011}, but this question cannot be 
clarified in this study.

However, we cannot exclude that there may be a small hysteresis for
the branch pieces to the left of minimum and to the right of minimum,
respectively (cf.~Fig.~\ref{fig:volfracandfluc_g}).  Clearly, if the
volume ensemble  completely described all structural degrees of
freedom and the probability, two states with the same $\beta$ and the
same $\phi$ would be identical and therefore $\sigma_\phi$ would also have to
be the same.  However, if the volume ensemble is only a good
approximation of the geometric aspects of interdependent
force-moment and volume ensembles (see, e.g.~\cite{Blumenfeld2012}),
deviations may occur.

\section{Limitations of the volume ensemble} \label{sec:limits}
\begin{figure}[h]
 \includegraphics[width=\columnwidth]{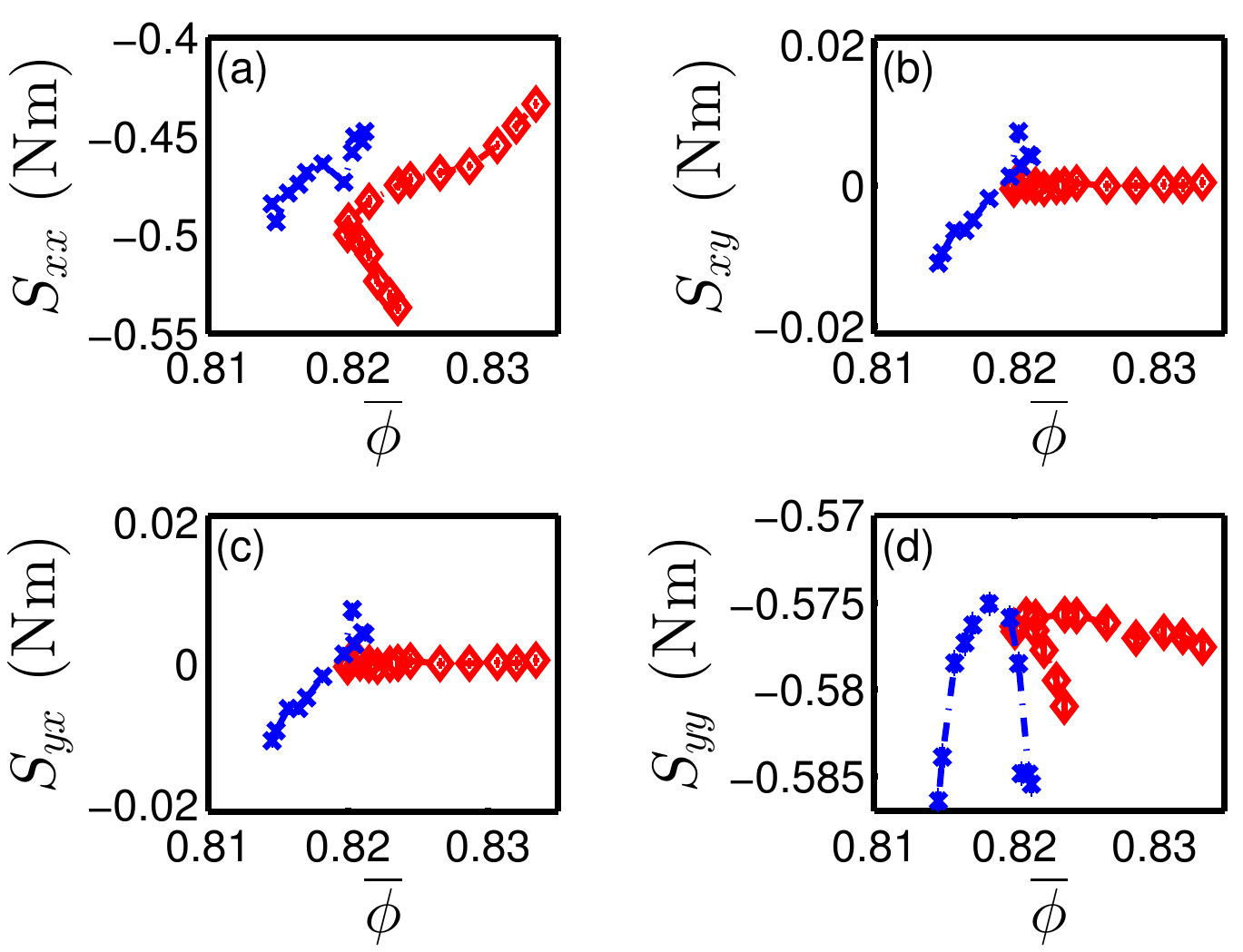}
 \caption{\label{fig:stress}(Color online) The components of the extensive stress
   tensor as function of the mean volume fraction for the ``negative
   g'' protocol (diamonds) and the ``rotating g'' protocol (crosses).}
\end{figure}

Whereas the volume ensemble appears to succeed in describing the
geometrical and structural degrees of freedom of a granular aggregate,
this is not the case for the stress state of the latter.  If the volume ensemble
entailed a complete description of a granular state and its
probability distribution, the mean stress of the system would have to
be a unique function of the inverse compactivity and therefore also a
unique function of the mean volume fraction.  We computed the mean
extensive stress (or force-moment tensor), defined as
\begin{equation}
 \vec{S}_{ij} =  \sum_{p} V^p \sigma^p_{ij} \label{eq:extStress}
\end{equation}
as a function of the volume fraction. Note that the volume density of
this tensor is the stress itself. Here $i$ and $j$ label Cartesian
coordinates. The sum runs over all particles, where $\sigma_{ij}^p$ is
the mean stress in particle $p$ and $V^p$ is the volume associated
with the particle.  The result is shown in Fig.~\ref{fig:stress}.  The
stress tensor is obviously not a unique function of the volume
fraction.

Neither the results from the different protocols nor the results from
the half-branches left to the minimum and right to the minimum,
respectively, of the ``negative g'' protocol fall on the same curve,
which is in agreement with similar findings on tapped granular matter
\cite{Pugnaloni2010}.  Two systems with almost identical particle
positions and orientations can be in very different stress states
which is not captured by the volume ensemble.  To describe the stress
state and the volume state together, probably the combined
volume-stress ensemble \cite{Blumenfeld2012} must be taken into
account. We remark that it is unclear so far wether or not the stress
states observed here are ``very different'' or even ``very similar''
because we do not know the size of the accessible phase space for the
extensive stress tensor.  Maybe the deviations in our systems which
have magnitude on the order of $0.01\, \text{Nm}$ are so small that it
is reasonable to assume in first approximation that the stress-moment
tensor is almost constant which would be a possible explanation for
the success of assuming a pure volume ensemble.

\section{Conclusions} \label{sec:conclusion}
 
We used two different protocols to excite a granular ensemble
periodically.  The inverse compactivity was determined as a function
of the mean volume fraction and we found that the relation between the
compactivity and the mean volume fraction is protocol independent. We
determined an expression for the logarithm of the partition function
and thus the thermodynamic potential which is the equivalent of the
free energy of classical statistics. This was done by using the ideal
quadron solution derived by Blumenfeld and Edwards
\cite{Blumenfeld2003} as a fitting function. Even though the ideal
quadron solution makes very rough assumptions, the resulting
description is in qualitative agreement with the findings from the
simulations. If the particle number is used as a fit parameter instead
of calculating the distribution with the true particle number, the
ideal quadron solution is also able to describe the results
quantitatively.

We found that all our simulation results related to structural
quantities are compatible with Edwards' theory and that the Edwards
theory describes the volume fluctuations very well,
independently of the excitation protocol. The usage of the Edwards
volume ensemble seems to be sufficient for the description of system
properties which are related to the geometrical arrangement of the
grains, but as might be expected from previous findings, it is not
sufficient to describe the stress states of the granular ensemble.


\begin{appendix}

\section{Overlapping histograms for Gaussian distributions}
\label{app:gauss}
If two volume samples were Gaussian distributed with mean values $V_1$
and $V_2$ and variances $\sigma_1^2$ and $\sigma_2^2$, the corresponding
$Q$-function defined in \eqref{eq:Qdefinition} would be
\cite{McNamara2009}:
\begin{equation}
  Q^g(V) = -\frac{(V-V_1)^2}{2 \sigma_1^2} + 
  \frac{(V-V_2)^2}{2 \sigma_2^2} + \ln \left( \frac{\sigma_2}{\sigma_1}  \right). 
  \label{eq:qgauss}
\end{equation}
This is a quadratic function of $V$, but in some  $V$ interval the
curvature of $Q^g$ may be very small and the parabolic function
\eqref{eq:qgauss} would then practically be indistinguishable from a linear
function. This happens in particular, if the variances $\sigma_1$ and
$\sigma_2$ are close to each other, i.e., for nearby compactivities.
If we define a function $A^g_{21}$ as the slope of \eqref{eq:qgauss}
midway between the maximum values of the two normal distributions,
\begin{equation}
  A^g_{21}:= \left. \frac{\D}{\D V} Q^g(V) \right|_{V=(V_1 
    + V_2)/2}\>,
\end{equation}
we obtain \cite{McNamara2009}:
\begin{equation}
  A^g_{21} = \frac{1}{2} \left(\frac{1}{\sigma_1^2} + \frac{1}{\sigma_2^2}\right)  
\left( V_1 -  V_2 \right). \label{eq:Adefinition}
\end{equation}
Identifying formally
\begin{equation}
 A^g_{21}=\beta(V_2)-\beta(V_1)\>, \label{eq:formalidentity}
\end{equation}
we find from \eqref{eq:Adefinition}:
\begin{equation}
 -\frac{V_2-V_1}{\beta(V_2)-\beta(V_1)} = 2 \left(\frac{1}{\sigma_1^2} + \frac{1}{\sigma_2^2}\right)^{-1}\>. \label{eq:alternative}
\end{equation}
If we assume that the variance is a unique function of the mean
volume, this equation is an approximation of \eqref{eq:sigmaVvonv}
which is the better the smaller the difference $|V_2-V_1|$.

We note that the formal identification \eqref{eq:formalidentity} is
strictly speaking inherently contradictory, which is easy to see if we
calculate $Q$ for a third sample with mean value $V_3$ and variance
$\sigma_3$ and the same reference sample in the denominator. From
\eqref{eq:formalidentity} $A_{31}-A_{21}=A_{32}$ follows, but if one
calculates the same quantity from \eqref{eq:Adefinition}, the result does not
agree.  (However, the identification is possible, if the three
variances are the same.)

Analogously, we can calculate the intercept $B^g_{21}$ of the tangent
which touches \eqref{eq:qgauss} at $V=(V1 + V2)/2$:
\begin{align}
  B^g_{21}&=\frac{1}{8} (V_2-V_1)^2 
  \left( \frac{1}{\sigma_2^2}-\frac{1}{\sigma_1^2} \right) 
  +\ln \left( \frac{\sigma_2}{\sigma_1} \right)
\nonumber \\
&\>\>+\frac14  \left(V_2^2-V_1^2\right)\left(\frac{1}{\sigma_1^2} + \frac{1}{\sigma_2^2}\right) \>.
\end{align}
By calculating the limit 
\begin{equation}
  \lim_{V_2 \rightarrow V_1} \frac{B^g_{21}}{A^g_{21}} = 
-{V}_1-\sigma_1\frac{\partial \sigma_1}{\partial V_1}
\end{equation}
we find that the term $B^g_{21}/A^g_{21}$ is an approximation for
\eqref{eq:meanVolume}, if we assume $\frac{\partial \sigma_1}{\partial
  V_1}$ to be negligibly small and formally identify
$B_{21}^g=\ln(Z_2)-\ln(Z_1)$. Again the approximation becomes better
as the difference between the volume $V_2$ and the reference volume
$V_1$ gets smaller.

Due to these similarities, it might occur that even if the relations
\eqref{eq:meanVolume}, \eqref{eq:sigmaVvonv} and
\eqref{eq:overlappinghistofit} are consistent, that within the limits
of data precision it is not possible to decide whether the reason of
this agreement is the correctness of Edwards' theory or the fact that the
data generated by a specific protocol happen to have a distribution
that is well approximated by a normal distribution. If the samples are
generated using the same protocol, one may expect that there is a
function $\sigma(V)$, but when different protocols are used, it would be
surprising, if both protocols led to the same relation, unless a
general principle, such as Edwards' theory, were at work.

On one hand, these similarities make it more difficult to verify
Edwards' theory, on the other hand, due to the central limit theorem,
we should expect that the distribution of Eq.~\eqref{eq:probfromvol}
becomes more and more Gaussian with increasing system size. Therefore,
it is not always true that the appearance of Gaussian distributions
signifies inapplicability of the overlapping histogram method in the
determination of the compactivity and of related quantities. Let us
briefly have a look at this. Rewriting Eq.~\eqref{eq:probfromvol} for
general $\beta$ and using the microcanonical result for the density of
states, we have
\begin{align}
  P(V,\beta,N)
  &= \frac{\Omega(V,N)}{Z(\beta,N)} e^{-\beta V}= \frac{e^{S(V,N)-\beta V}}{Z(\beta,N)} \>.
\end{align}
As in standard statistical mechanics, we can then argue that for
\emph{large} systems this distribution has a sharp peak at the mean
value of the volume and we may expand the entropy about this average,
neglecting terms that are of higher than quadratic order:
\begin{align}
  S(V,N)-\beta V &\approx S(\bar{V},N) + \beta(\bar V)
  \left(V-\bar{V}\right)
  \nonumber\\
  &\>\>+\frac12 \frac{\partial^2 S}{\partial V^2}\Big\vert_{\bar V}
  \left(V-\bar{V}\right)^2 -\beta V 
\>.
 \label{eq:S_expans}
\end{align}
In order for the expansion to be about the maximum of the distribution,
we must require the linear order term to vanish, i.e., we have $\beta(\bar
V)=\beta$, meaning equivalence of the microcanonical and the canonical
compactivity de\-finitions. Identifying the inverse of $-\partial^2
S/\partial V^2$ with the variance, our distribution takes the form
\begin{align}
  P(V,\beta,N)&= \frac{e^{S(\bar V,N)-\beta \bar V -
    \left(V-\bar{V}\right)^2/2\sigma^2}}{Z(\beta,N)}\>.
\end{align}
Evaluating this at two different compactivities and taking the
logarithm of the ratio, we find (denoting the mean volumes by $V_1$ and
$V_2$ again)
\begin{align}
  Q(V) &= \ln \frac{P(V,\beta_1,N)}{P(V,\beta_2,N)} =
  -\frac{(V-V_1)^2}{2 \sigma_1^2} + \frac{(V-V_2)^2}{2 \sigma_2^2}
\nonumber\\
  &\>\>-\beta_1 V_1 + S(V_1,N) + \beta_2 V_2 - S(V_2,N)
\nonumber\\
&\>\> + \ln\frac{Z(\beta_2,N)}{Z(\beta_1,N)}\>, 
\end{align}
which is nothing but \eqref{eq:qgauss} with an explicit expression for
$\ln (\sigma_2/\sigma_1)$. But we have derived this as an approximation
to the distribution \eqref{eq:probfromvol} from which we obtain
Eq.~\eqref{eq:overlappinghistofit} for $Q$. If we substitute $\beta_2$
for $\beta_0$ in that equation, we see that the following
(non-trivial) approximation holds in sufficiently large systems (close
to the ``thermodynamic limit''), as long as the distributions overlap
significantly (which of course becomes less likely with increasing
system size):
\begin{align}
  &-(\beta_1-\beta_2) V \approx
  -\frac{(V-V_1)^2}{2 \sigma_1^2} + \frac{(V-V_2)^2}{2 \sigma_2^2}
  \nonumber\\
  &\>\>-\beta_1 V_1 + S(V_1,N) + \beta_2 V_2 - S(V_2,N)\>.
\end{align}
Hence, the sum on the right-hand side that is quadratic in $V$ is a
good approximation to the sum on the left-hand side that is linear in
$V$. As we have shown by this small calculation, the
overlapping-histogram method will give, for such a system, the correct
linear dependence on $V$, despite the fact that the central part of
the distribution is well approximated by a Gaussian. In simulations,
this behavior might be distinguished from Gaussian distributions not
having the statistical mechanical origin postulated by the Edwards
theory through verification that the tails of the simulated
distributions, i.e., their behavior for $V$ values, where the
quadratic approximation \eqref{eq:S_expans} breaks down, are not
Gaussian.  This was done for our simulations via the chi-square test
mentioned in Sec.~\ref{sec:anwendung}.


\section{Kernel density estimation} \label{append:kde} 
\begin{figure}[t!]
 \includegraphics[width=\columnwidth]{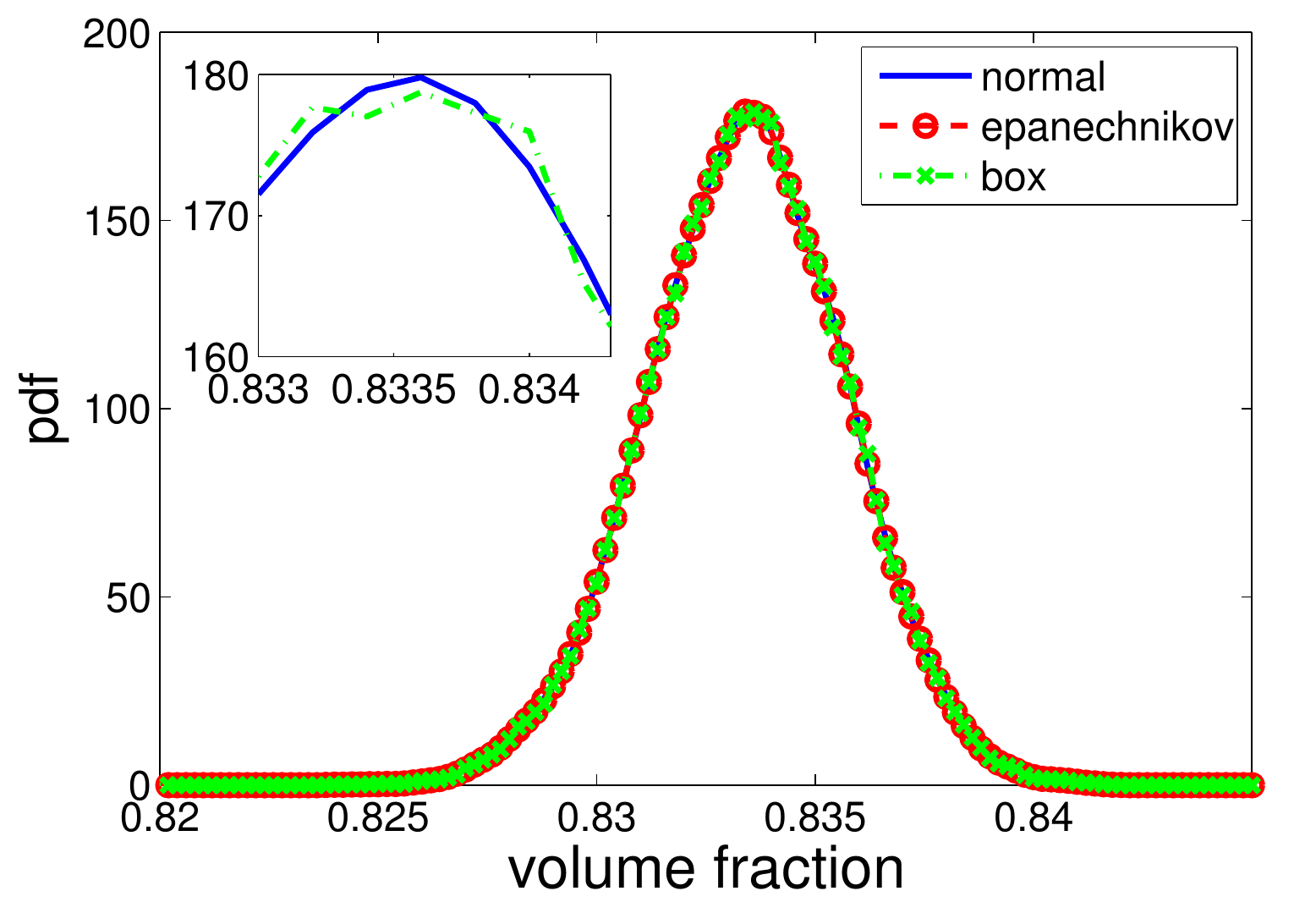}
 \caption{\label{fig:comparekernel}(Color onlie) Comparison of the results of
   kernel density estimation using different kernels.  The shown data
   was obtained from the volume fraction time series of the ``negative
   g'' protocol with the tapping parameter $g_p=8$, for other
   parameters we found the same conclusion. The results due to the
   Epanechnikov kernel and the normal kernel are indistinguishable, the
   box kernel results in a little bit more irregular estimations. The
   inset shows the normal and the box kernel in the central region of
   the distribution.}
\end{figure}

A method to
determine a continuous probability density from a data sample is the
kernel density estimation method (KDE)
\cite{Parzen1962,Rosenblatt1956}.  If $x_1,x_2,...,x_n$ are sampled
data, the kernel density estimation of the probability density $P(x)$
at the point $x$ is defined as
 \begin{equation}
   P(x) = \frac{1}{n h} \sum_{i=1}^n K\left( \frac{x-x_i}{h} \right)\>, 
 \end{equation}
 where $K(x)$ is the kernel which must be a non negative function that
 satisfies
\begin{equation}
 \int_{-\infty}^{\infty} \text{d}x~K(x) =1\>.
\end{equation}
and $h$ is a smoothing parameter called the bandwidth.
Possible kernels are for example  the normal kernel
\begin{equation}
   K(x)=\frac{1}{2 \pi} \exp(-x^2/2)\>,
\end{equation}
the Cauchy kernel
\begin{equation}
  K=\frac{1}{\pi(1+x^2)}\>,
\end{equation}  
  the Epanechnikov kernel
\begin{equation}
    K(t)=\begin{cases} \frac{3}{4}(1-x^2) & \text{if}~x \in [-1;1] \\
 0 & \text{elsewhere} \end{cases}\>,
\end{equation}
or even the rectangular kernel:
\begin{equation}
  K(t)=\begin{cases} 1 & \text{if}~x \in [-1/2;1/2] \\ 0 & \text{elsewhere} \end{cases}\>.
\end{equation}
The latter is equivalent to a histogram with bin width $h$. While it
can be shown that the Epanechnikov kernel is optimal in the sense that
it minimizes the mean squared error between the estimated and the real
probability distributions, we used a normal kernel since it allows to
make a good estimation of the optimal bandwidth $h$.  In general, the
optimal bandwidth can only be calculated if one knows the underlying
probability density, but this density is unknown.  In practice,
therefore, Silverman's rule of thumb is commonly used. Under the
assumption that the underlying probability distribution is Gaussian
and if a Gaussian kernel is used, the optimal bandwidth is
\begin{equation}
 \left( \frac{4 \sigma^5}{3n} \right)^{\frac{1}{5}} \simeq 1.06 \sigma n^{-1/5}\>,
\end{equation}
where $\sigma$ is the standard derivation of the sample. It turns out
that this bandwidth is also a reasonable choice in practical
situations, if the underlying distribution is not Gaussian. We 
note that for small data sets the choice of the kernel may have a
significant influence on the quality of the fit. However, if the data
sample becomes big enough, all kernels leads to almost the same
results except for the far tail of the distribution, where no data
points are available. In this region, the kernel itself specifies the
 decay of the distribution. As it is shown exemplarily in Figure
\ref{fig:comparekernel}, the choice of the kernel is not crucial for
our data samples. The results obtained with the optimal Epanechnikov
Kernel and the results achieved with the normal kernel are practically
indistinguishable.  The box kernel which is equivalent to a shifted
histogram is a little bit more more irregular.  We preferred the
normal kernel, in order to avoid an \emph{ad hoc} choice of 
the kernel bandwidth.

\end{appendix}

%

\end{document}